\documentclass[english,longauth]{aa}
\usepackage[T1]{fontenc}
\usepackage[latin1]{inputenc}
\usepackage{rotating}
\usepackage{textcomp}
\usepackage{graphicx}
\IfFileExists{url.sty}{\usepackage{url}}
                      {\newcommand{\url}{\texttt}}
\usepackage[authoryear]{natbib}

\makeatletter

\usepackage{color}
\usepackage{hyperref}

\usepackage{units}
\usepackage{natbib}

\bibpunct{(}{)}{;}{a}{}{,}

\usepackage{babel}
\makeatother

\begin{document}

\author{A.~Nigl\inst{1}
\and W.D.~Apel\inst{2}
\and J.C.~Arteaga\inst{3}
\and T.~Asch\inst{4}
\and J.~Auffenberg\inst{5}
\and F.~Badea\inst{2}
\and L.~B\"ahren\inst{6}
\and K.~Bekk\inst{2}
\and M.~Bertaina\inst{7}
\and P.L.~Biermann\inst{8}
\and J.~Bl\"umer\inst{2,3}
\and H.~Bozdog\inst{2}
\and I.M.~Brancus\inst{9}
\and M.~Br\"uggemann\inst{10}
\and P.~Buchholz\inst{10}
\and S.~Buitink\inst{1}
\and H.~Butcher\inst{6}
\and E.~Cantoni\inst{7}
\and A.~Chiavassa\inst{7}
\and F.~Cossavella\inst{3}
\and K.~Daumiller\inst{2}
\and V.~de Souza\inst{3}
\and F.~Di Pierro\inst{7}
\and P.~Doll\inst{2}
\and R.~Engel\inst{2}
\and H.~Falcke\inst{1,6}
\and H.~Gemmeke\inst{4}
\and P.L.~Ghia\inst{11}
\and R.~Glasstetter\inst{5}
\and C.~Grupen\inst{10}
\and A.~Haungs\inst{2}
\and D.~Heck\inst{2}
\and J.R.~H\"orandel\inst{1}
\and A.~Horneffer\inst{1}
\and T.~Huege\inst{2}
\and P.G.~Isar\inst{2}
\and K.-H.~Kampert\inst{5}
\and D.~Kickelbick\inst{10}
\and Y.~Kolotaev\inst{10}
\and O.~Kr\"omer\inst{4}
\and J.~Kuijpers\inst{1}
\and S.~Lafebre\inst{1}
\and P.~\L{}uczak\inst{12}
\and M.~Manewald\inst{4}
\and H.J.~Mathes\inst{2}
\and H.J.~Mayer\inst{2}
\and C.~Meurer\inst{2}
\and B.~Mitrica\inst{9}
\and C.~Morello\inst{11}
\and G.~Navarra\inst{7}
\and S.~Nehls\inst{2}
\and J.~Oehlschl\"ager\inst{2}
\and S.~Ostapchenko\inst{2}
\and S.~Over\inst{10}
\and M.~Petcu\inst{9}
\and T.~Pierog\inst{2}
\and J.~Rautenberg\inst{5}
\and H.~Rebel\inst{2}
\and M.~Roth\inst{2}
\and A.~Saftoiu\inst{9}
\and H.~Schieler\inst{2}
\and A.~Schmidt\inst{4}
\and F.~Schr\"oder\inst{2}
\and O.~Sima\inst{13}
\and K.~Singh\inst{1}
\and M.~St\"umpert\inst{3}
\and G.~Toma\inst{9}
\and G.C.~Trinchero\inst{11}
\and H.~Ulrich\inst{2}
\and J.~van~Buren\inst{2}
\and W.~Walkowiak\inst{10}
\and A.~Weindl\inst{2}
\and J.~Wochele\inst{2}
\and J.~Zabierowski\inst{12}
\and J.A.~Zensus\inst{8}
}

\institute{ 
Department of Astrophysics, IMAPP, Radboud University Nijmegen, P.O. Box 9010, 6500 GL Nijmegen, The~Netherlands\\
\email{anigl@astro.ru.nl}\\
  \and 
Institut\ f\"ur Kernphysik, Forschungszentrum Karlsruhe, 76021~Karlsruhe, Germany\\
  \and 
Institut f\"ur Experimentelle Kernphysik, Universit\"at Karlsruhe (TH), 76021~Karlsruhe, Germany\\
  \and 
Institut f\"ur Prozessverarb. und Elektr., Forschungszentrum Karlsruhe, 76021~Karlsruhe, Germany\\
  \and 
Fachbereich Physik, Universit\"at Wuppertal, 42097~Wuppertal, Germany \\
  \and 
ASTRON, 7990 AA Dwingeloo, The Netherlands \\
  \and 
Dipartimento di Fisica Generale dell'Universit{\`a}, 10125~Torino, Italy\\
  \and 
Max-Planck-Institut f\"ur Radioastronomie, 53010~Bonn, Germany \\
  \and 
 National Institute of Physics and Nuclear Engineering, 7690~Bucharest, Romania\\
  \and 
Fachbereich Physik, Universit\"at Siegen, 57068~Siegen, Germany \\
  \and 
Istituto di Fisica dello Spazio Interplanetario, INAF, 10133~Torino, Italy \\
  \and 
Soltan Institute for Nuclear Studies, 90950~Lodz, Poland\\
  \and 
Physics Department, Bucharest University, Bucharest-Magurele, P.O. Box MG-11, RO-077125, Romania
}

\title{Frequency spectra of cosmic ray air shower radio emission measured with LOPES}

\abstract
{}
{We wish to study the spectral dependence of the radio emission from cosmic-ray air showers around $100~\unit{PeV}$ $(10^{17}~\unit{eV})$.}
{We observe short radio pulses in a broad frequency band with the dipole-interferometer LOPES (LOFAR Prototype Station), which is triggered by a particle detector array named Karlsruhe Shower Core and Array Detector (KASCADE). LOFAR is the Low Frequency Array. For this analysis, 23 strong air shower events are selected using parameters from KASCADE. The radio data are digitally beam-formed before the spectra are determined by sub-band filtering and fast Fourier transformation.}
{The resulting electric field spectra fall off to higher frequencies. An average electric field spectrum is fitted with an exponential $E_{\nu}=K\cdot\mbox{exp}(\nu/\unit{MHz}/\beta)$ and $\beta=-0.017\pm0.004$, or alternatively, with a power law $\epsilon_{\nu}=K\cdot\nu^{\alpha}$ and a spectral index of $\alpha=-1\pm0.2$. The spectral slope obtained is not consistent within uncertainties and it is slightly steeper than the slope obtained from Monte Carlo simulations based on air showers simulated with CORSIKA (Cosmic Ray Simulations for KASCADE). For the analyzed sample of LOPES events, we do not find any significant dependence of the spectral slope on the electric field amplitude, the azimuth angle, the zenith angle, the curvature radius, nor on the average distance of the antennae from the shower core position. But one of the strongest events was measured during thunderstorm activity in the vicinity of LOPES and shows the longest pulse length measured of $110~\unit{ns}$ and a spectral slope of $\alpha=-3.6$.}
{We show with two different methods that frequency spectra from air shower radio emission can be reconstructed on event-by-event basis, with only two dozen dipole antennae simultaneously over a broad range of frequencies. According to the obtained spectral slopes, the maximum power is emitted below 40 MHz. Furthermore, the decrease in power to higher frequencies indicates a loss in coherence determined by the shower disc thickness. We conclude that a broader bandwidth, larger collecting area, and longer baselines, as will be provided by LOFAR, are necessary to further investigate the relation of the coherence, pulse length, and spectral slope of cosmic ray air showers.}

\authorrunning{Nigl et al.}
\titlerunning{CR electric field spectrum with LOPES}
\offprints{A. Nigl}
\mail{anigl@astro.ru.nl}
\date{Received 7 December 2008 / Accepted 17 June 2008}

\maketitle

\section{Introduction}

Cosmic rays are particles or nuclei constantly bombarding the Earth's
atmosphere with a large spread in energy. The origin of cosmic rays
with the highest energies is still unknown and therefore subject of
intensive research (\citealt{pao07}). These particles, mostly protons,
initiate an extensive shower of secondary particles traveling with
almost the speed of light through the air. The charged particles in
the air shower produce electromagnetic emission relativistically beamed
in the forward direction. This work probes the spectral dependence
of the electromagnetic field produced in these air showers to investigate
the radio emission mechanism. Understanding the emission mechanism
and its dependence on frequency and distance to the shower core will
be an important step to infer the primary particle species of the
cosmic ray.

The radio emission of charged particles in air showers was first observed
in 1964 by \citet{jelley65} at 44~MHz and by \citealt{allan66}
at 60~MHz. Those early studies found strongly pulsed radio emission
from air showers detected with radio antennae triggered by particle
detectors. A dependence of the radio emission's polarization on the
geomagnetic field favored a geomagnetic emission mechanism (see, e.g.,
\citealt[ 1969]{allan67}; \citealt{sun75}). The first quantitative
radio frequency spectra were obtained from a few simultaneous narrowband
observations below 100~MHz by, e.g., \citet{spencer69}, \citet{allan70},
and \citet{prah71}. In this earlier work, spectra decreasing in frequency
were obtained with spectral indices $\alpha$ between $-2$~and~$-1$.
However, radio measurements were abandoned in the late 1970s because
of strong radio frequency interference in the signal, difficulties
with the interpretation of the measurements, and because of the success
of alternative observing techniques. An excellent review of the early
work can be found in \citet{allan71}.

Interest in the radio technique was revived by the LOFAR Prototype
Station (LOPES) group (\citealt{falcke05}) and by the Cosmic ray
Detection Array with Logarithmic ElectroMagnetic Antennas (CODALEMA)
group (\citealt{codalema06}), both using digital data acquisition
and powerful computers for data analysis on digitized radio signal
for the first time. LOFAR is the Low Frequency Array. LOPES measured
a linear dependence of the electric field amplitude of the radio emission
in extensive cosmic-ray air showers on the energy of the primary particle.
This result was an important step in proving the radio emission process
to be coherent.

The following two radio emission mechanisms are believed to mainly
contribute to the production of electromagnetic radiation in air showers.
First, the geomagnetic mechanism generates beamed synchrotron emission
through relativistic electron-positron pairs in the Earth's magnetic
field, which is believed to be the dominant mechanism (\citealt{allan67};
\citealt{allan69}). This ansatz led to the description of coherent
geosynchrotron emission at meter wavelengths (\citealt{falcke03}).
Second, relativistic particles emit Cherenkov radiation as a kind
of photonic shockwave (\citealt[1965]{askaryan62}; \citealt{kahn66}).
However, the Cherenkov emission mechanism is more efficient in denser
media than air, like water and ice. Additionally, radio emission in
air showers was claimed to be dependent on the geoelectric field,
and it was found that strong electric fields in thunderstorm clouds
can amplify the measured radio amplitudes (\citealt{mandolesi74};
\citealt{gurevich04}; \citealt{buitink07}).

When a cosmic ray initiates an extensive air shower, a thin shower
pancake of particles is formed growing to about a few hundred meters
in diameter. The pancake has a thickness of a few meters, due to particle
attenuation in the atmosphere and relativistic beaming at high gamma
factors. Close to the axis of the cosmic-ray air shower, the particle
pancake is thinnest and thus the pulse profile received on the axis
is expected to be shortest. The radio pulse is expected to broaden
with distance from the shower core, mainly caused by geometric-delays
in the arrival time of longitudinal shower evolution stages (\citealt{huege07}).
The spectral shape of the radio emission is mainly determined by the
distribution of electrons and positrons in the shower pancake, as
they emit synchrotron radiation. According to Monte Carlo simulations,
the spectrum starts with coherent emission at wavelengths larger than
the shower pancake thickness, and it falls off exponentially to higher
frequencies, where coherence will be lost when the wavelength becomes
comparable with the shower thickness (\citealt{huege05aap}).

A cosmic-ray radio spectrum is particularly interesting to probe the
coherence of the geosynchrotron emission and to study the width of
the radio pancake, which determines the length of the detected pulse.
This is important for future instruments, such as LOFAR, which will
provide even larger bandwidths and baselines than LOPES.

We present electric field spectra determined from data recorded by
LOPES with 30 antennae dipoles. The instruments LOPES and KASCADE
are described in Sect. \ref{sec:cr_instrument}. The selection and
data analysis of radio events are explained in Sect. \ref{sec:cr_eventselection}
and in Sect. \ref{sec:cr_datareduction}, respectively. The obtained
spectra are shown and discussed in Sect. \ref{sec:cr_fieldspec}.

\section{Instrument\label{sec:cr_instrument}}

\begin{figure}
\begin{centering}
\includegraphics[bb=0bp 0bp 500bp 500bp,width=1\columnwidth]{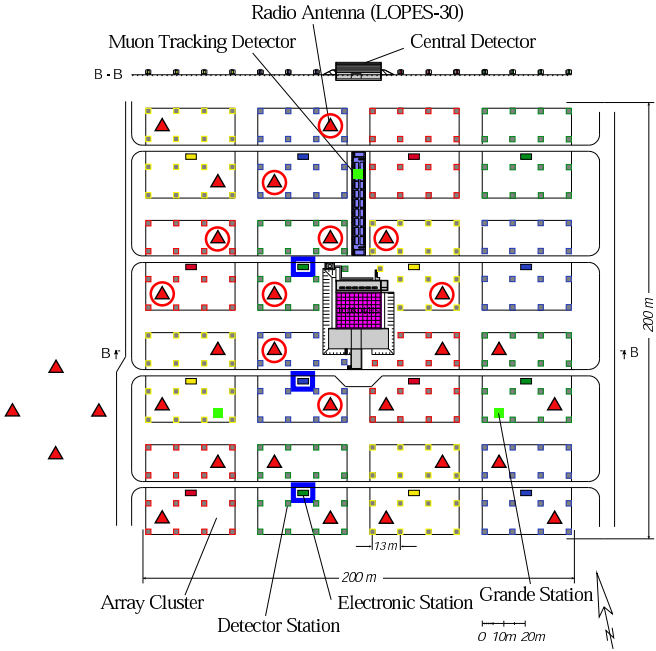} 
\par\end{centering}

\caption{\label{fig:cr_layout}Layout of LOPES inside the KASCADE array (\citealt{horneffer06}).
The 252 small squares indicate the KASCADE particle detectors. The
triangles show the positions of the LOPES antennae, as they were positioned
for this work, and the circles highlight the first stage of LOPES
with 10 operational antennae. The three rectangles mark the electronic
stations that house the LOPES electronics. Each of these stations
is collecting data from ten antennae elements. The top station holds
the master clock module and a board for the reception of the KASCADE
timestamp. Among the KASCADE detectors are three KASCADE-Grande (\citealt{navarra04})
detectors (squares), which are spaced on a bigger grid to cover a
larger area and thus higher primary particle energies.}

\end{figure}

The LOFAR prototype station LOPES (\url{www.lopes-project.org}, \citealt{horneffer04,falcke05})
has been optimized for the detection of cosmic rays with energies
around $10^{17}~\unit{eV}$. LOPES is placed in between particle detectors
of the experiment KASCADE (\url{www-ik.fzk.de/KASCADE_home.html},
\citealt{antoni03}) in Karlsruhe. The data for this work were taken
with LOPES operating 30 early LOFAR prototype dipole antennae. KASCADE
is the Karlsruhe Shower Core and Array Detector located at the Forschungszentrum
Karlsruhe, Germany. It is made up of $16\times16$ particle detector
stations placed on a $200\times200~\unit{m^{2}}$ rectangular grid,
with 13~m spacing, a central hadron calorimeter, and a muon tracking
detector. A layout of LOPES and KASCADE on the site of the Forschungszentrum
Karlsruhe is shown in Fig. \ref{fig:cr_layout}.

KASCADE provides LOPES with a cosmic-ray trigger as well as particle
shower parameters, like muon number, electron number, direction of
origin, and position of the shower center (\citealt{antoni03}). These
parameters have been used to select cosmic-ray events for LOPES post-processing.

The LOPES hardware digitizes the radio signals measured by the dipole
antenna wires with a rate of 80 Megasamples per second and in a dynamic
range of 12 bits. The frequency band used for cosmic-ray detection
ranges from 43~MHz to 74~MHz, since the bandpass filter attenuates
the received signal significantly outside these limits. The data is
stored on custom-made PC boards with a 1~GB ring buffer for each
dipole. The readout time of a 0.8~ms cosmic-ray event takes about
1.5~s and the system is not sensitive to a new trigger for that period.
The data taken with LOPES are matched with the KASCADE data and sent
to storage units for further off-line analysis. LOPES was triggered
with a large-event-trigger from KASCADE, which requires a detection
in 10 out of the 16 detector clusters of KASCADE.

\subsection{Radio pulse broadening}

LOPES measures the pulse at antenna level, thus at different positions
and distances from the shower core at the same time. Radio frequency
interference (RFI) is rejected by down-weighting of narrow spectral
lines that lie three sigmas above the average spectrum in two iterations.
The signals of all antenna dipoles are combined by beam-forming to
increase sensitivity and to further suppress RFI. The 30 antenna positions
in the LOPES layout result in an average distance of the antennae
from any possible shower core position within KASCADE of $\bar{d}=75~\unit{m}$.

Furthermore, the signal detected by the LOPES dipoles is bandpass
filtered within the limits from 40~MHz to 80~MHz. In principle,
the filtered bandwidth of 40~MHz broadens an infinitely short pulse
in time to $25~\unit{ns}$ ($\Delta t\sim1/\Delta\nu$) and changes
the spectral slope outside the band due to attenuation. The impulse
response of a first LOPES antenna fitted with a Gaussian resulted
in a width of $57~\unit{ns}$ (\citealt{horneffer06}). Thus, additional
broadening must be caused by the phase-characteristic of the LOPES
hardware, but the spectral slope is not changed within the LOPES band.
Therefore, the measured pulse lengths are upper limits and the minimum
detectable width is close to $25~\unit{ns}$.

\subsection{Gain calibration\label{sub:crspec_gaincal}}

\begin{figure}
\begin{centering}
\includegraphics[width=1\columnwidth]{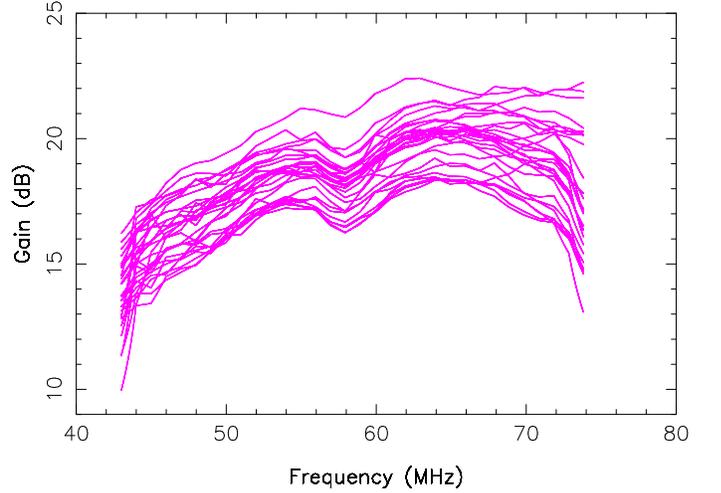} 
\par\end{centering}

\caption{\label{fig:cr_el_gain} Electronics gain factors are plotted for all
30 LOPES antennae. The factors are for the correction of the received
voltage measured in several calibration campaigns at KASCADE in the
band from 43~MHz to 74~MHz.}

\end{figure}

The inverted V-shaped dipole antennae of LOPES have a direction dependent
antenna gain and the data acquisition hardware as a whole has a frequency
dependent electronics gain. The electronics gain was determined for
each antenna in several dedicated calibration campaigns by measuring
a well-defined signal from a reference antenna mounted on a crane
overhead LOPES (\citealt{nehls07}, Fig. \ref{fig:cr_el_gain}).

\begin{figure}
\begin{centering}
\includegraphics[width=0.75\columnwidth]{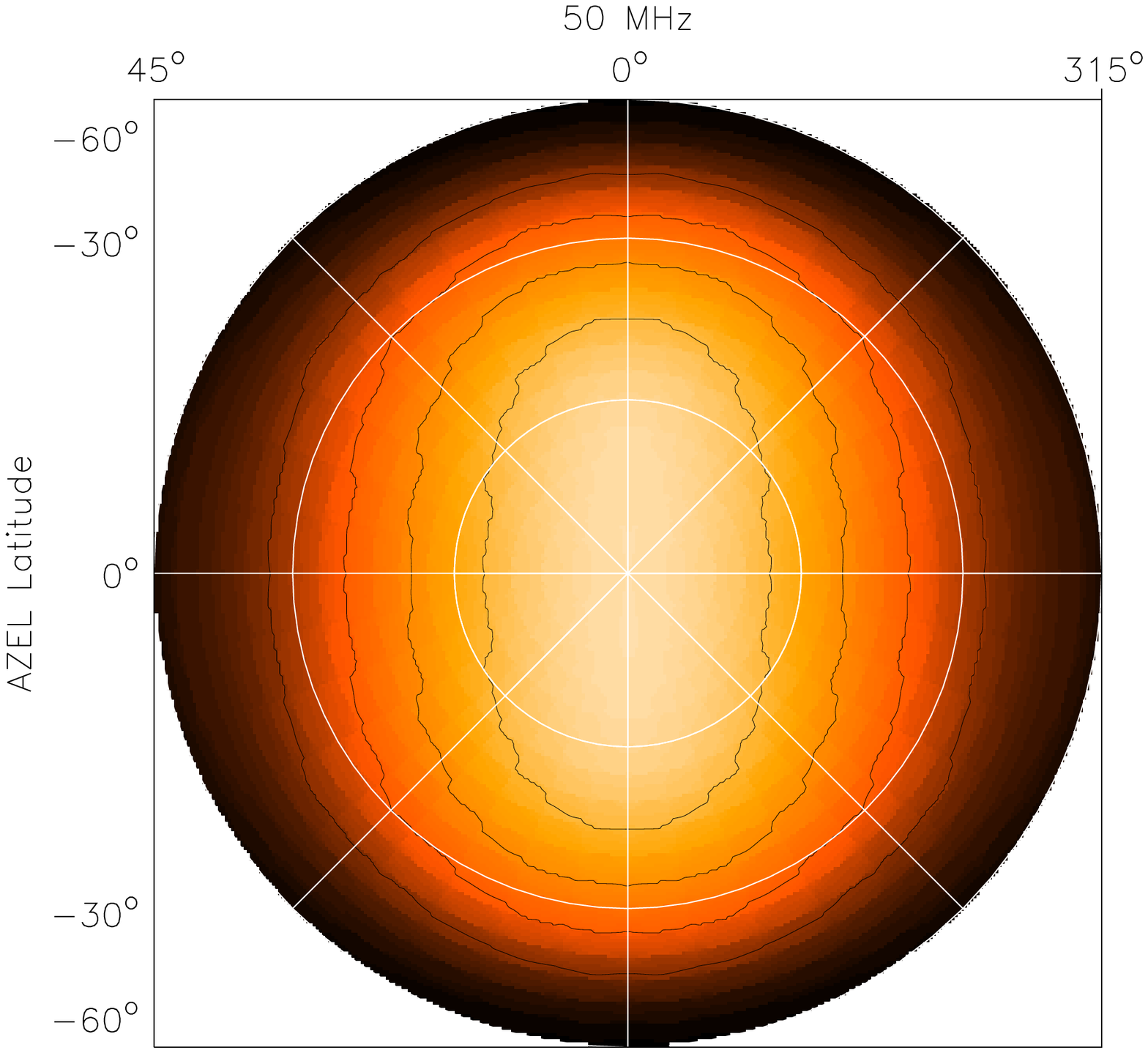} 
\par\end{centering}

\begin{centering}
\includegraphics[width=0.75\columnwidth]{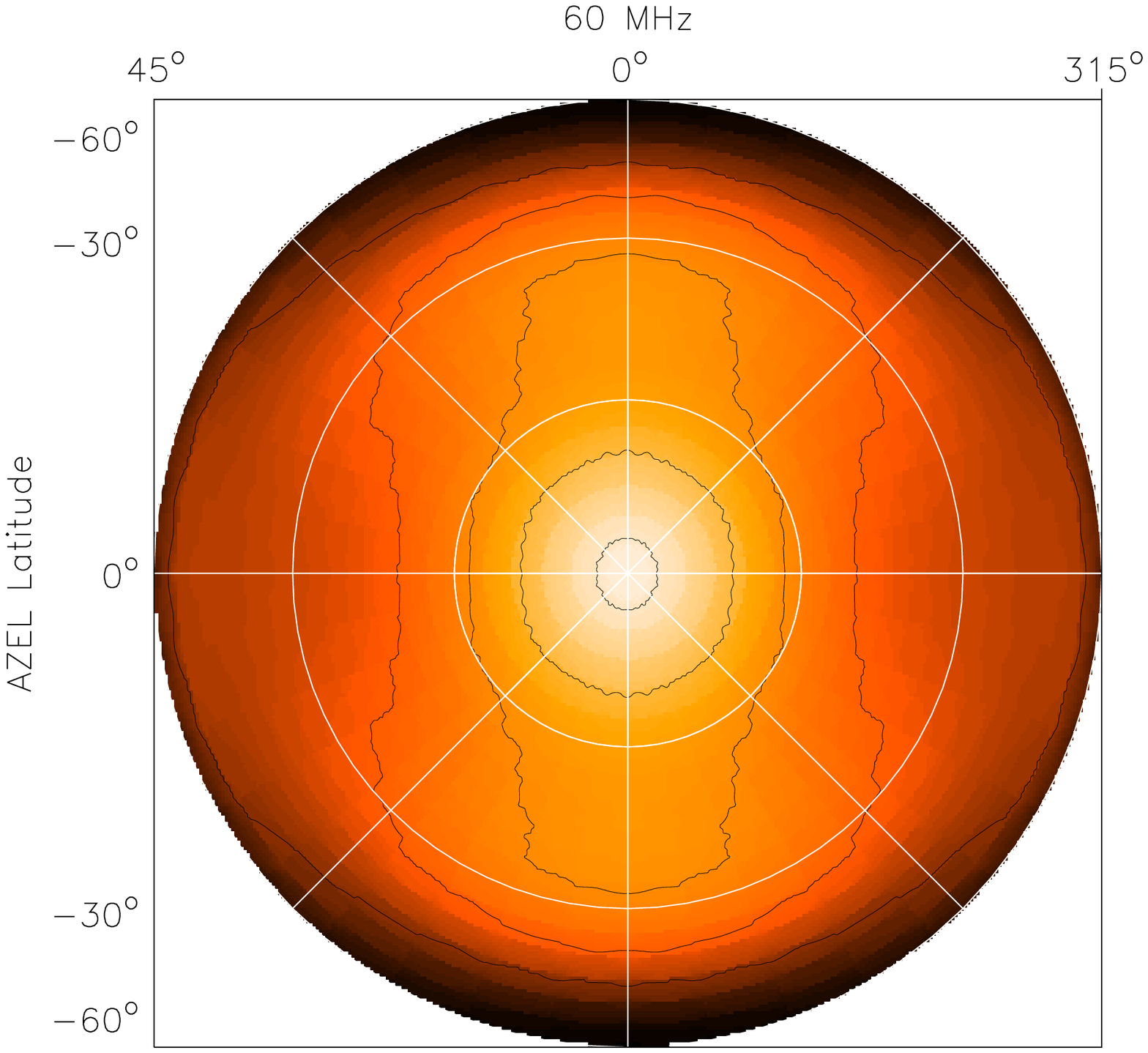}
\par\end{centering}

\begin{centering}
\includegraphics[width=0.75\columnwidth]{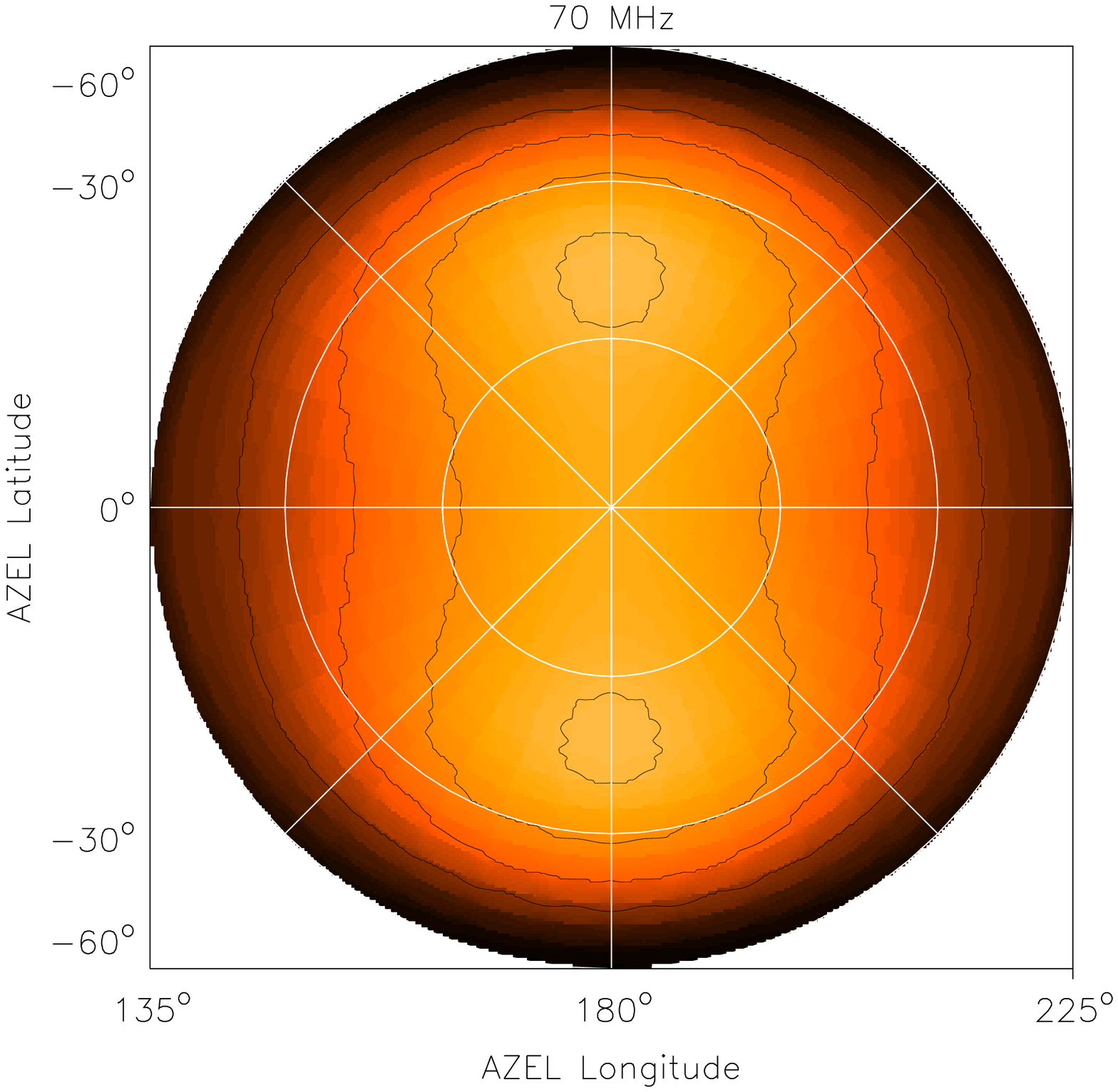}
\par\end{centering}

\caption{\label{fig:cr_gainpattern}Simulated gain patterns for a single LOPES
dipole antenna, for all directions in azimuth $\mathbf{\mathnormal{\phi}}$
and elevation $\mathbf{\mathnormal{\theta}}$, at frequencies 50~MHz
(top), 60~MHz (middle) and 70~MHz (bottom), for a bandwidth of 1~MHz.
The zenith lies in the center of each plot and the white circles increase
in steps of $30\unit{^{\circ}}$ in elevation down to the horizon
at the plot edges. The antenna dipole is lying horizontally in the
plots. The gain ranges from its minimum in black (gain~=~0) to the
maximum gain in white (gain = 5). The dark contours are plotted in
steps of 20\%. Where five contours are visible (middle plot), they
go from 20\% gain to 100\% gain.}

\end{figure}

The direction and frequency dependent antenna gain calibration values
$G_{(\phi,\theta,\nu)}$ for a LOPES antenna element, mounted on a
metal pedestal, were calculated with a simulation model by \citet{arts05}.
Here, $\phi$ is the azimuth angle and $\theta$ is the elevation
angle. The pedestal and the geometry of the LOPES dipoles make the
gain pattern frequency dependent. At the lower end of the LOPES band,
the contour-lines of equal gain in the pattern are oval-shaped and
the dipoles have their maximum sensitivity to the zenith with the
major axis of the ellipse being perpendicular to the dipole, lying
in East-West direction. With increasing frequency, the sensitivity
maximum gets circular in shape concentrating around the zenith (in
the middle of the band) and then the direction of maximum gain weakens,
splits, and moves down in elevation at azimuth angles perpendicular
to the dipole (see Fig. \ref{fig:cr_gainpattern}).

The output of the analog-digital converter (ADC) can be converted
to units of field strength per unit bandwidth, using both sets of
gains (\citealt{horneffer06}):\begin{eqnarray}
\left\langle \mbox{{\bf $\varepsilon$}}_{\nu}\right\rangle  & = & \frac{|{\bf E}|}{\Delta\nu}\nonumber \\
 & = & \frac{1}{\Delta\nu}\sqrt{\frac{4\pi\nu^{2}\mu_{0}}{G_{(\phi,\theta,\nu)}c}P_{ant}}\label{eq:cr_gain_cal}\\
 & = & \frac{1}{\Delta\nu}\sqrt{\frac{4\pi\nu^{2}\mu_{0}}{G_{(\phi,\theta,\nu)}c}K_{\nu}\frac{V_{ADC}^{2}}{R_{ADC}}}.\nonumber \end{eqnarray}
Here, ${\bf E}$ is the electric field from the air shower measured
at the antenna; $\Delta\nu$ is the filtered bandwidth of the signal;
$\nu$ is the observing frequency; $G_{(\phi,\theta,\nu)}$ is the
direction and frequency dependent gain of the antenna (Fig. \ref{fig:cr_gainpattern});
$P_{ant}$ is the power received by the antenna; $K_{\nu}$ is the
frequency dependent correction factor for the electronics; as plotted
in Fig. \ref{fig:cr_el_gain} for each antenna; $V_{ADC}$ is the
voltage digitized by the ADC; $R_{ADC}$ is the input impedance of
the ADC; $\mu_{0}$ is the electromagnetic permeability; and $c$
the speed of light.

For earlier LOPES results, the directional gain was averaged in frequency
and applied to the beam-formed time-series. For this work, it was
applied in the frequency domain to each individual frequency bin (see
Sect. \ref{sub:cr_pipeline}).

\section{Event selection\label{sec:cr_eventselection}}

For this analysis, we studied events of LOPES using the 30 antennae
setup from 2005-11-16 to 2006-11-30. In this period, LOPES was recording
in a single polarization configuration, all dipoles were parallel
and most sensitive to radiation from the East-West direction. About
one million events were triggered and recorded within this period.

Events were pre-selected on parameters provided by KASCADE and by
estimating the electric field strength:\begin{eqnarray}
\epsilon_{{\rm est}} & = & (53\pm4.9)\left[\frac{{\rm \mu V}}{{\rm m\, MHz}}\right]\nonumber \\
 &  & \times\left((1.09\pm0.017)-\cos\vartheta\right)\label{eq:cr_horneffer_muon_icrc}\\
 &  & \times\exp\left(\frac{{\rm -R_{SA}}}{{\rm (221\pm62)\, m}}\right)\left(\frac{{\rm N_{\mu}}}{10^{6}}\right)^{(0.99\pm0.04)}.\nonumber \end{eqnarray}

This formula is obtained from fits to all LOPES events detected with
the standard analysis software (\citealt{horneffer07}). Here, $N_{µ}$
is the muon number at distances between 40~m and 200~m, which is
provided by KASCADE, $\vartheta$ is the geomagnetic angle (calculated
using the shower direction reconstructed by KASCADE), and $R_{SA}$
is the average distance of the antennae from the shower axis.

We limited the distance of the shower core from the KASCADE center
to 91~m to make sure that the shower core is inside the KASCADE array
so that the shower reconstruction by KASCADE provides high accuracy.
For this work, 71 radio events were pre-selected with a lower limit
of $4~\unit{µV~m^{-1}~MHz^{-1}}$ on the electric field strength obtained
with Eq. \ref{eq:cr_horneffer_muon_icrc}.

For the spectral analysis, 23 events of the 71 events were selected
with a lower limit on the signal-to-noise ratio (SNR) of the E-field
amplitude and the E-field root-mean-square (RMS) of 5 for each spectral
bin. For these 23 events, a radio spectrum was generated with the
two methods described in Sect. \ref{sec:cr_datareduction}. The estimated
primary energy for these processed radio events lies mainly in the
range from $10^{17}~\unit{eV}$ to $10^{18}~\unit{eV}$.

\section{Data reduction\label{sec:cr_datareduction}}

The standard analysis software is a modular software package and it
was developed for the data reduction of LOPES radio events. We use
this software to pre-process the data for this work. The software
allows us to view the data at any step in the desired sequence of
processing (\citealt{horneffer06}). After calibration, digital filtering,
and beam-forming of the selected events, we performed the spectral
analysis according to the following processing pipeline.

\subsection{Processing pipeline\label{sub:cr_pipeline}}

\begin{figure}
\begin{centering}
\includegraphics[width=0.9\columnwidth]{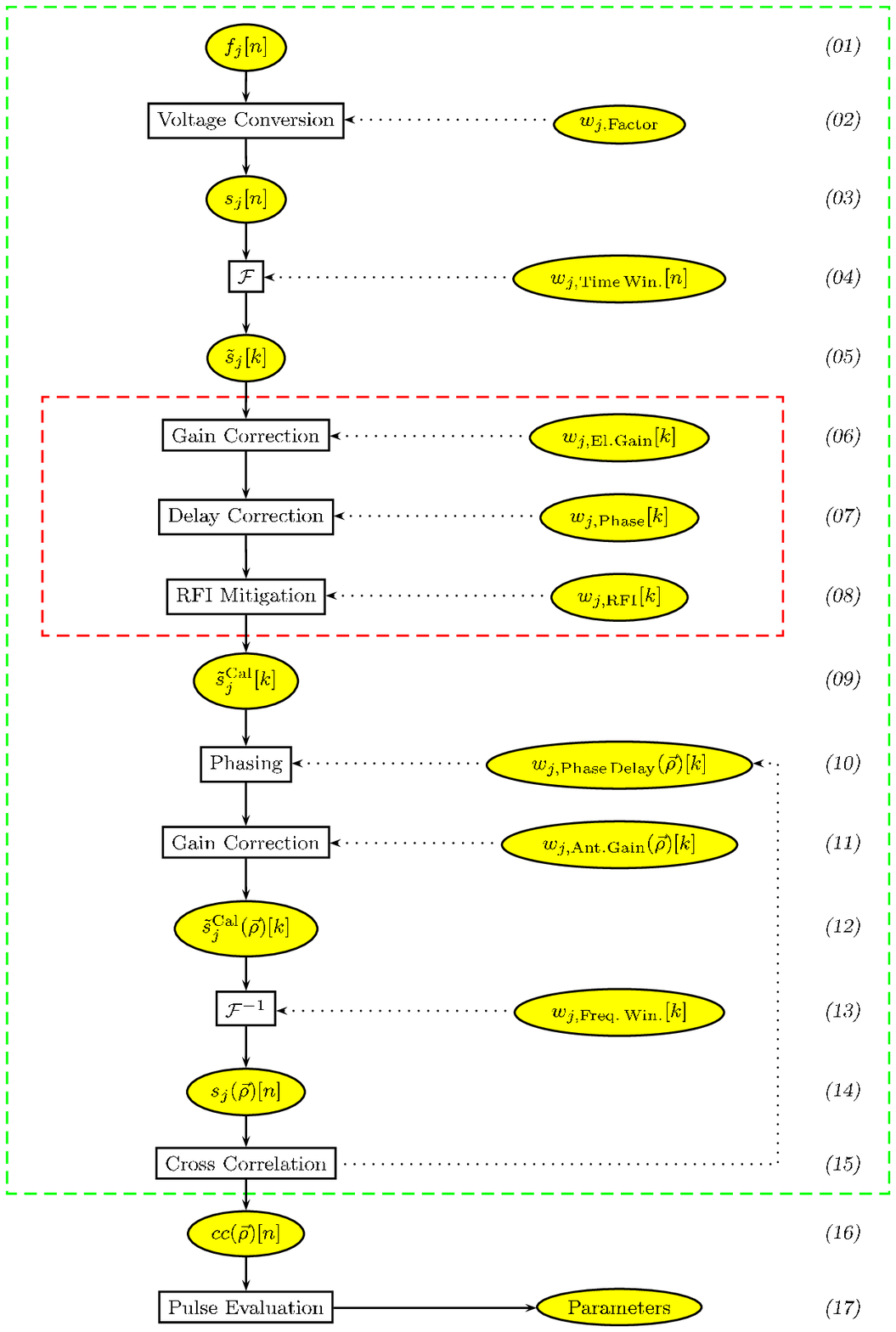} 
\par\end{centering}

\caption{\label{fig:cr_pipeline}This flowchart of the cosmic-ray processing
pipeline starts from the raw data $f_{j}[n]$, where $n$ is the index
for the ADC samples per antenna $j$. The signal converted from ADC
counts to voltage is denoted as $s$. The Fourier transformation is
denoted as $\mathcal{F}$ and the inverse Fourier transformation as
$\mathcal{F}^{-1}$. The number of the frequency bins after Fourier
transformation is indicated with $k$. Directional dependence of the
signal and the weights is indicated by the vector $\protect\overrightarrow{\rho}$.
The outer dashed line encloses the processing per individual dipole
and the inner dashed line encloses the signal calibration steps in
the frequency domain. A more detailed description of the steps in
the processing pipeline can be found in Sect. \ref{sub:cr_pipeline}.}

\end{figure}

Two methods have been used to produce the electric field spectra.
First, the determination of the electric field amplitude of the cosmic-ray
radio pulse detected in the antennae cross-correlation beam (see Sect.
\ref{sub:cr_ccbeam}) was filtered in several adjacent frequency bins
of the LOPES band from 43 MHz to 74 MHz. Second, a Fourier transformation
was applied to a few tens of samples around the full-band radio peak
detected in the simple antennae beam (see Sect. \ref{sub:cr_fbeam}).

The processing structure for the production of the cosmic-ray electric
field spectra is shown in Fig. \ref{fig:cr_pipeline}. The digitized
signal is converted from ADC counts (01) to voltage (02). A Hanning
window is applied to the time-series data (03) to reduce leakage in
the successive Fast Fourier Transform (FFT) (04). In the frequency
domain, weights for the frequency dependent electronics gain (06),
weights for correcting delays introduced by the instrument hardware
(07) and weights to reduce RFI (08) are applied to the frequency bins
(05).

The resulting calibrated antenna signals (09) (as measured at the
antenna feeds) are multiplied with another set of complex weights
to phase them in the direction of the cosmic ray (10). The direction
is provided by the KASCADE experiment and the standard analysis software
attempts to refine it by fitting a maximum to the cosmic-ray pulse
emission in a radio skymap (see Sect. \ref{sub:cr_dir}).

The phased antenna spectra (12) are corrected by the directional gain
of the antennae (11) and they are split in 16 sub-bands of 2.5~MHz
applying modified Hanning window functions (13) as described in Sect.
\ref{sub:cr_freq_win_fct}. After inverse Fourier transformation,
the digitally-filtered and calibrated antenna time-series (14) are
cross-correlated (15) to form the cc-beam, which is described in Sect.
\ref{sub:cr_ccbeam}. For some events it became necessary to review
the antenna selection and the beam forming result, and to reject antennae
that do not contribute to a coherent pulse, which is indicated by
the recursive arrow in Fig. \ref{fig:cr_pipeline}.

From the resulting cc-beam (16) the background noise is subtracted
(see Sect. \ref{sub:cr_background_noise}) and the amplitude of the
electric field strength of the pulse (17) is determined by averaging
32 samples ($400~\unit{ns}$) around the pulse to obtain the electric
field spectrum.

\begin{figure}
\begin{centering}
\includegraphics[width=1\columnwidth]{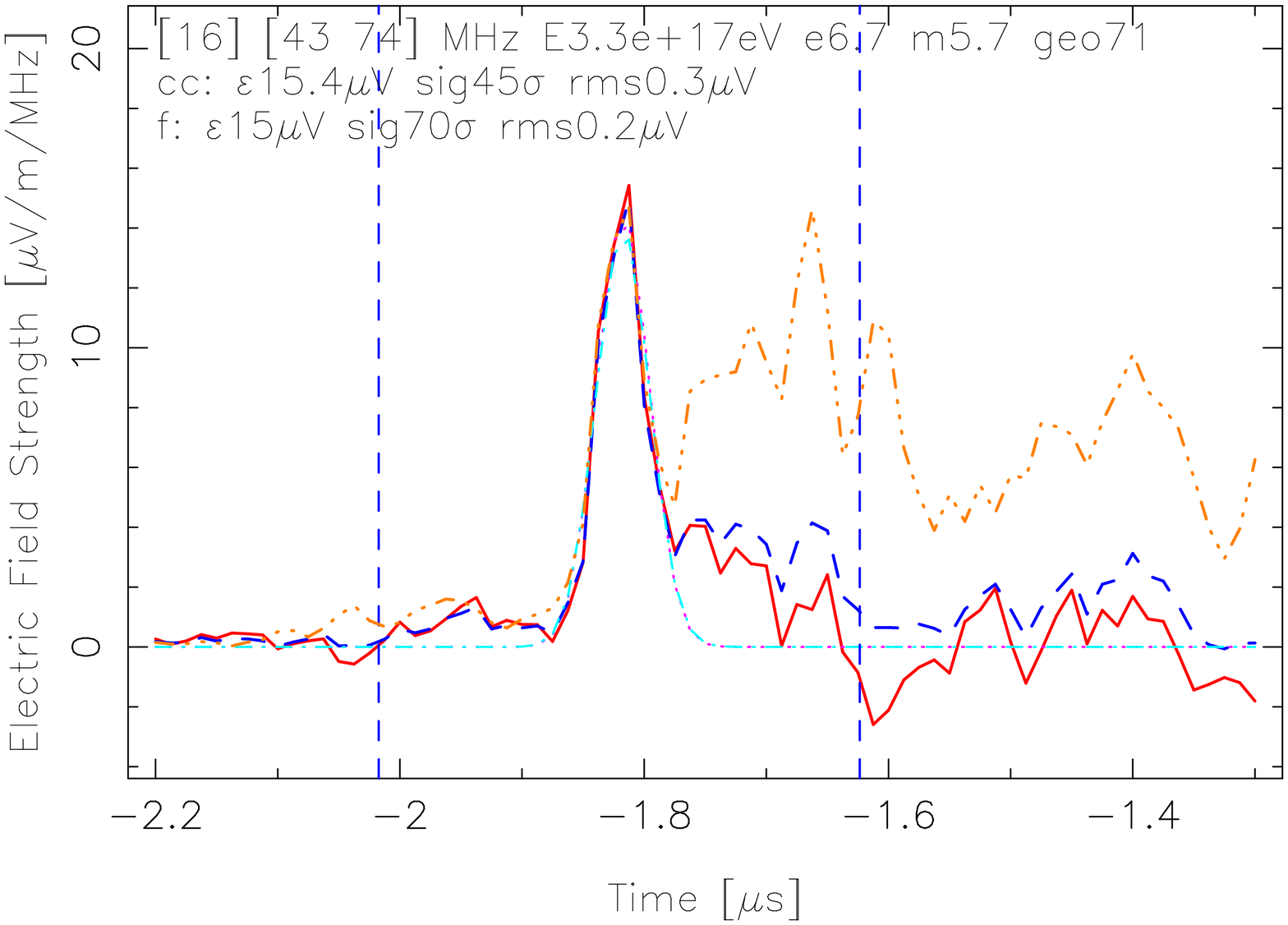}
\par\end{centering}

\caption{\label{fig:cr_fullbeam}Comparison of full-band cross-correlation
beam (cc-beam, solid line), full-band field strength beam (f-beam,
dashed), and full-band power beam (p-beam, dash dotted) for a characteristic
strong event. Additionally, Gauss-fits to the cc-beam and f-beam are
plotted (thin dashed line and thin dotted line). The vertical lines
(dashed) indicate the time window of the f-beam for which the FFT
was applied.}

\vspace{20pt}

\begin{centering}
\includegraphics[width=1\columnwidth]{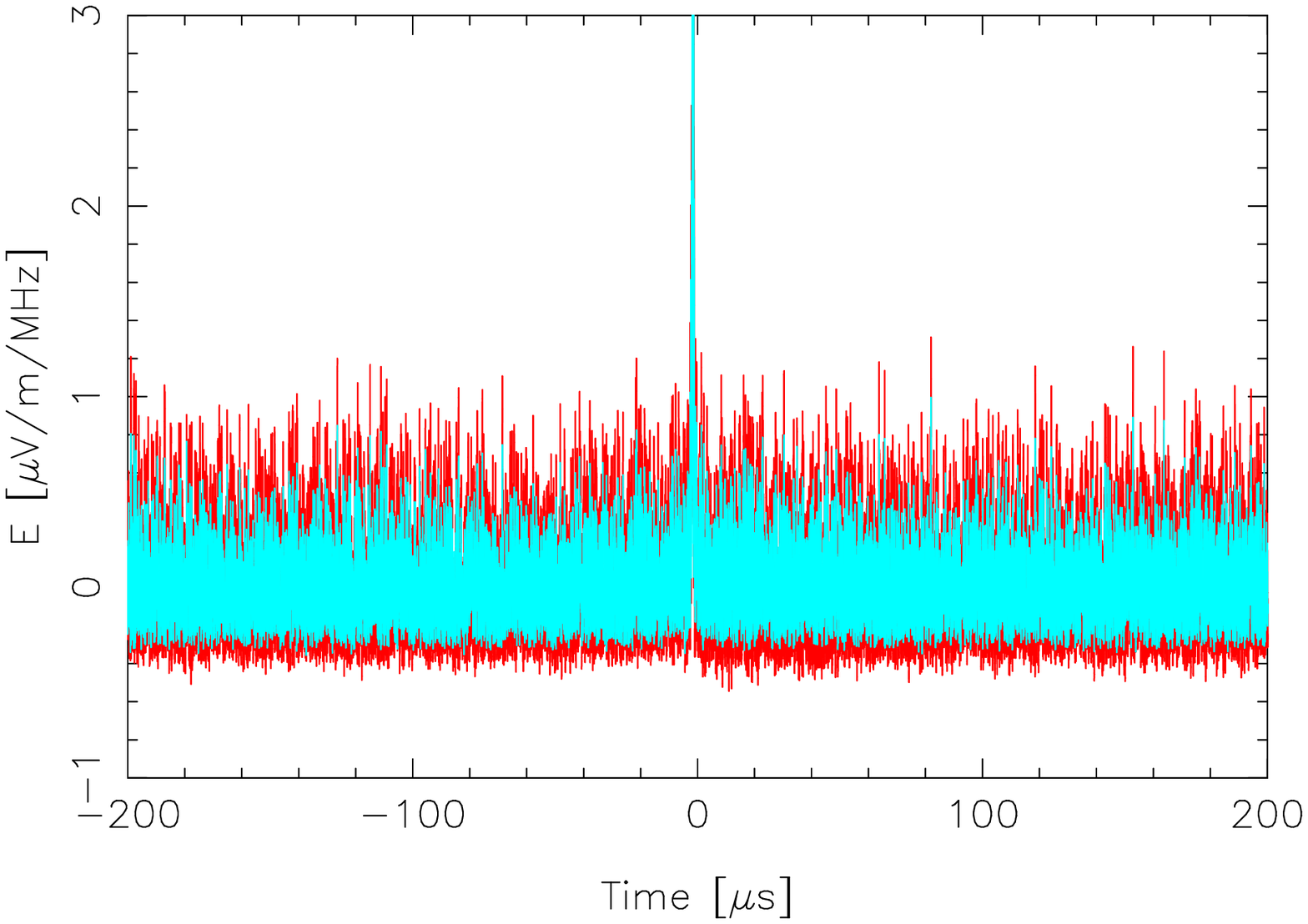}
\par\end{centering}

\caption{\label{fig:cr_fullbeamnoise}Comparison of the noise of the cross-correlation
beam (cc-beam: large RMS) and the noise of the field strength beam
(f-beam: small RMS) of a LOPES event for the period of 0.2~ms before
and after the trigger. The peak slightly left of center of the time-window
is the air shower pulse.}

\end{figure}

For the second method, the phased full-band antenna spectra (12) are
inverse Fourier transformed back to time. The resulting filtered and
calibrated antenna time-series are averaged to form the simple f-beam
(see Sect. \ref{sub:cr_fbeam}). The cc-beam was not used here since
it calculates the absolute value of the electric field in the beam,
whereas for the FFT-method, the full phase information is needed and
thus the f-beam was used instead (see Sect. \ref{sub:cr_fbeam}).
The f-beam is also reduced by its background noise level and a modified
Hanning window is applied. Subsequently, the pulse found in the f-beam
is Fourier transformed back to the frequency domain to obtain the
electric field spectrum. For the FFT, 32 samples ($400~\unit{ns}$)
around the air shower pulse (see Fig. \ref{fig:cr_fullbeam}) were
chosen to minimize a contribution to the signal from the particle
detectors and to obtain a spectral resolution of 2.5~MHz matching
the resolution of the first method. The values in the top left corner
of the plot read: in the 1st line: the event number, the LOPES frequency
window from $43~\unit{MHz}$ to $74~\unit{MHz}$, the estimated primary
energy of $3.3\times10^{17}~\unit{eV}$, the total electron number
of $10^{6.7}$, the truncated muon number of $10^{5.7}$, and the
geomagnetic angle of $71\unit{^{\circ}}$; in the 2nd line: the electric
field amplitude of the cc-beam (cc) of $15.4~\unit{\mu V~m^{-1}MHz^{-1}}$,
the full-band peak SNR of $45\sigma$ and the root-mean-square (RMS)
of $0.3~\unit{\mu V~m^{-1}MHz^{-1}}$; and in the 3rd line: the electric
field amplitude of the f-beam (f) of $15~\unit{\mu V~m^{-1}MHz^{-1}}$,
the full-band peak SNR of $70\sigma$ and the RMS of $0.2~\unit{\mu V~m^{-1}MHz^{-1}}$.
The RMS of both full-band beams was determined on 0.2~ms of data
well before and after the air shower pulse (see Fig. \ref{fig:cr_fullbeamnoise}).

\subsection{Frequency window function\label{sub:cr_freq_win_fct}}

\begin{figure}
\begin{centering}
\includegraphics[width=1\columnwidth]{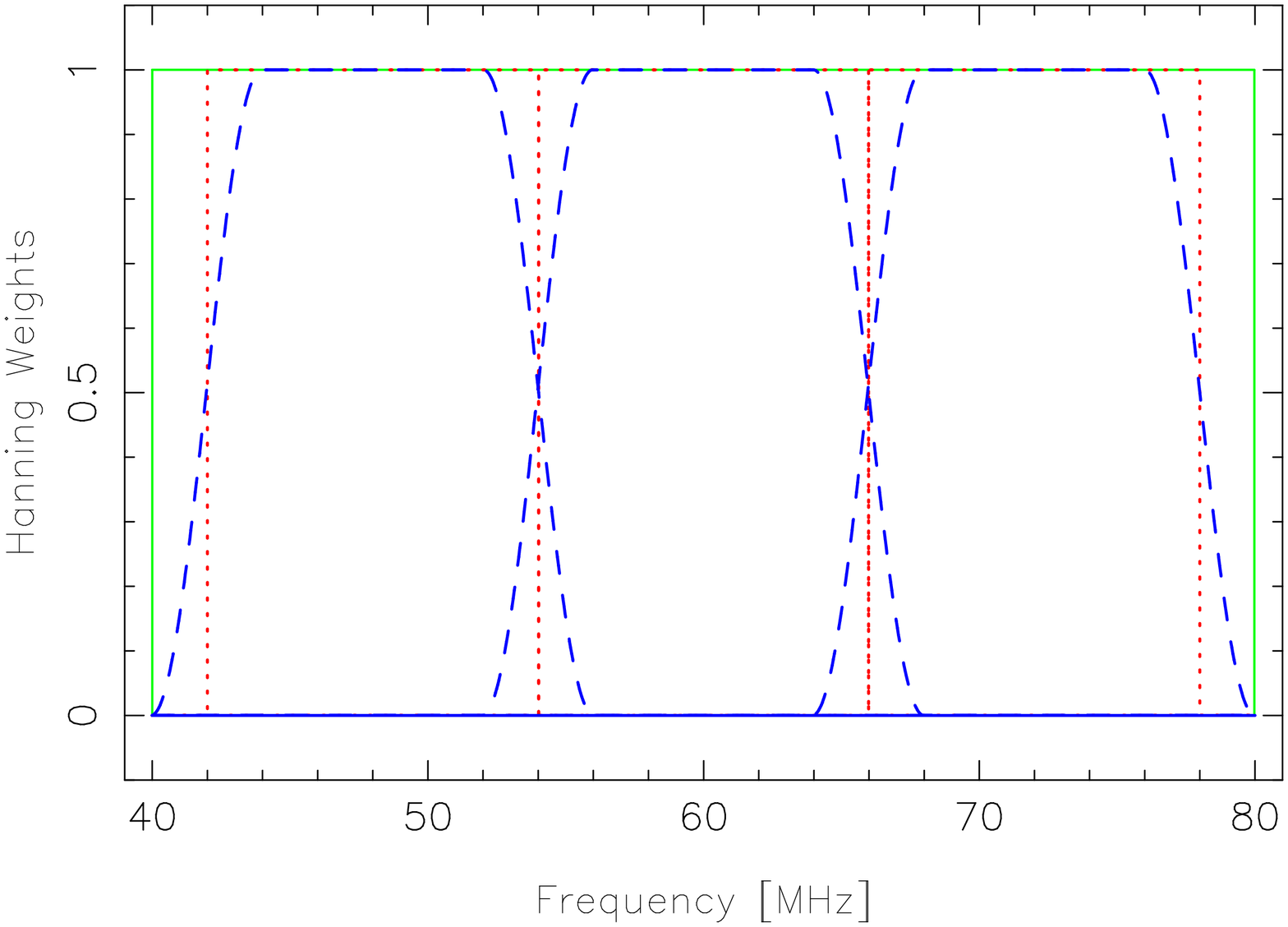} 
\par\end{centering}

\caption{\label{fig:cr_hanning} Modified Hanning windows (dashed) and rectangular
windows (dotted) for three sub-bands of the frequency band from 40~MHz
to 80~MHz.}

\end{figure}

For the selection of the frequency sub-bands, a filter function has
been applied. The simplest window is a rectangular window. However,
the sharp edges introduce leakage. For less sharp edges, a modified
Hanning window is used. The Hanning function applies only to the beginning
quarter and to the end quarter of the window. The Hanning window is
basically a cosine-function and it has to be applied to the frequency
band around the center frequency increased by half the period of the
cosine of the Hanning window to preserve the integral of the rectangular
window (see Fig. \ref{fig:cr_hanning}).

\subsection{Core position and beam direction\label{sub:cr_dir}}

The core position and the direction used for beam-forming is provided
by the KASCADE experiment. The accuracy of the core position and shower
direction of KASCADE for the selected showers of high energy are $1~\unit{m}$
and $0.1~\unit{^{\circ}}$, respectively, valid up to a zenith angle
of $42\unit{^{\circ}}$ (\citealt{antoni03}). The KASCADE direction
is optimized by finding the position of maximum emission in a 4D radio
image produced with the LOPES data. The axes of the four dimensions
are time, azimuth angle, elevation angle, and curvature radius of
an assumed spherical wavefront. The separation from the KASCADE direction
for the selected events increases slightly with zenith angle and ranges
up to $2.9\unit{^{\circ}}$ with an average of $(1.0±0.5)\unit{^{\circ}}$
for zenith angles smaller than $75\unit{^{\circ}}$, which is consistent
with an earlier result of $(0.8±0.4)\unit{^{\circ}}$ in \citet{falcke05},
where only events with a maximum zenith angle of $60\unit{^{\circ}}$
were analyzed (one sigma statistical uncertainties).

\subsection{Field strength beam\label{sub:cr_fbeam}}

The field strength beam (f-beam) is the average of multiple antennae
signals being shifted by geometrical delays for the beam direction.
The delayed antennae signals are added and then divided by the number
of antennae:\begin{eqnarray}
S_{f}(t) & = & \frac{1}{N}\sum_{i=1}^{N}f_{i}(t-\tau_{i}).\label{eq:cr_fbeam}\end{eqnarray}
Here, $N$ is the number of antennae, $f$ is the single antenna field
strength time-series, and $\tau$ the signal arrival time delay for
the beam direction.

\subsection{Cross-correlation beam\label{sub:cr_ccbeam}}

The cross-correlation beam (cc-beam) is defined as the real part of
a standard cross-correlation of multiple antennae signals, shifted
by one set of signal delays for a certain direction. The data from
each unique antenna pair are multiplied with each other, the resulting
values are averaged, and then the square root is taken while preserving
the sign:\begin{eqnarray}
S_{cc}(t) & = & ±\sqrt{\left|\frac{1}{N_{pairs}}\sum_{i=1}^{N-1}\sum_{j>i}^{N}f_{i}(t-\tau_{i})\, f_{j}(t-\tau_{j})\right|}.\label{eq:cr_ccbeam}\end{eqnarray}

Here, $N$ is the number of antennae, $N_{pairs}=N(N-1)/2$ is the
number of unique pairs of antennae, $f$ is the single antenna field
strength time-series, and $\tau$ the signal arrival time delay for
the beam direction. The advantage of the cc-beam, compared to the
simple field strength beam, is that only the coherent signal is preserved
within the beam and RFI, which is only seen by a single antenna is
more strongly suppressed, since auto-correlation terms are not taken
into account.

\subsection{Power beam\label{sub:cr_pbeam}}

The power beam (p-beam) is the average of all antenna auto-correlations:\begin{eqnarray}
S_{p}(t) & = & \sqrt{\frac{1}{N}\sum_{i=1}^{N}f_{i}^{2}(t)}.\label{eq:cr_pbeam}\end{eqnarray}

Here, $N$ is the number of antennae and $f$ is the single antenna
field strength time-series. The power beam is sensitive to the total
power received by all the antennae from all directions, independent
of the coherence of the signal.

\subsection{Background noise\label{sub:cr_background_noise}}

The system noise in a LOPES beam toward the zenith was determined
by averaging background spectra of 664 arbitrary chosen events recorded
over a period of one year (11/2005 - 11/2006). For these background
spectra, short pulses in time and narrowband signals in frequency
were attenuated by down-weighting of single samples in time and single
bins in frequency containing power exceeding one sigma above the average
in two iterations. The result in units of electric field strength
is plotted in Fig. \ref{fig:cr_noiseFspec} and in units of noise
temperature is plotted in Fig. \ref{fig:cr_noiseTspec}. The power
in the frequency bins at 52.5~MHz and 67.5~MHz of the field spectrum
contained strong variations, which suggests that not all RFI could
be rejected before the calculation of the spectrum. The spectral field
strength was obtained by averaging the result of the FFT applied to
the filtered time-series of each event and the noise temperature was
calculated on the average spectral power of all events. The latter
plot also shows the galactic noise interpolated with a model for 45~MHz
to 408~MHz: $T_{\nu}=32~\unit{K}\times\left(\nu/408~\unit{MHz}\right)^{-2.5}$
(\citealt{falcke03}). The galactic noise is lying only slightly below
the system noise of LOPES. Furthermore, the power during two periods
of 48 hours, separated by half a year, showed a modulation with 20\%
of the received signal in accordance with the galactic plane crossing
the zenith. Thus, a LOPES beam is sensitive to the galactic noise,
but still dominated by the system noise. The values outside the band
from 45~MHz to 72.5~MHz are strongly attenuated by the bandpass
filter.

We calculated the level of the background noise for the cc-beam and
for the f-beam on an event-by-event basis on two blocks of $0.2~\unit{ms}$
in the first and second half of each event. On average in time $n_{cc}=(-0.13\pm0.06)~\unit{µV~m^{-1}MHz^{-1}}$
and $n_{f}=(0.6\pm0.2)~\unit{µV~m^{-1}MHz^{-1}}$ (one sigma statistical
uncertainty).

The parts of the event files for the noise-calculation leave out the
event edges, which are affected by the Hanning window. Also, they
leave out the cosmic-ray pulse and possible emission from the KASCADE
particle detectors.

In the LOPES data, the signal of the particle detectors is trailing
the radio signal of the air shower, and the KASCADE trigger arrives
about $1.8~\unit{\mu s}$ later at the LOPES electronics than the
radio pulse. The peak of the KASCADE emission arrives shortly after
the real air shower pulse, between $1.7~\unit{\mu s}$ and $1.5~\unit{\mu s}$
before the trigger arrives at the LOPES electronics and it is less
than $1~\unit{\mu s}$ wide. The emission from the particle detectors
is received by each antenna from several directions. Therefore, the
detector noise does not add coherently and it is attenuated in the
beam-forming process.

The root-mean-square (RMS) of the background of the cc-beam was determined
for each frequency bin on the same parts of each event as the noise
level above. The RMS of the f-beam of each event was determined as
an average of 50 spectra calculated on blocks of 32 samples, which
were chosen away from the cosmic-ray pulse.

\subsection{Slope correction}

In Fig. \ref{fig:crspec_NindVSze}, the slope of the 664 noise spectra
recorded over the earlier mentioned period of one year is plotted
as a function of zenith angle. The plot shows a steepening of the
slopes of events with higher zenith angles. The variation of the sky
noise is averaged out over the period of one year and it is not observed
to change with zenith angle. A small contribution might come from
remaining RFI increasing in signal strength to the horizon. However,
the main contribution is introduced by deviation of the frequency
dependent antenna gain calibration from the real antenna gain. This
dependence affecting the slopes of the air shower spectra was corrected
by dividing each spectra by a normalized correction spectrum:\begin{eqnarray}
\hat{\epsilon}_{\nu} & = & \nu^{\alpha_{f}-\alpha_{r}}/\overline{\nu^{\alpha_{f}-\alpha_{r}}}.\label{eq:crspec_slopecor}\end{eqnarray}

\begin{figure}
\begin{centering}
\includegraphics[width=0.8\columnwidth]{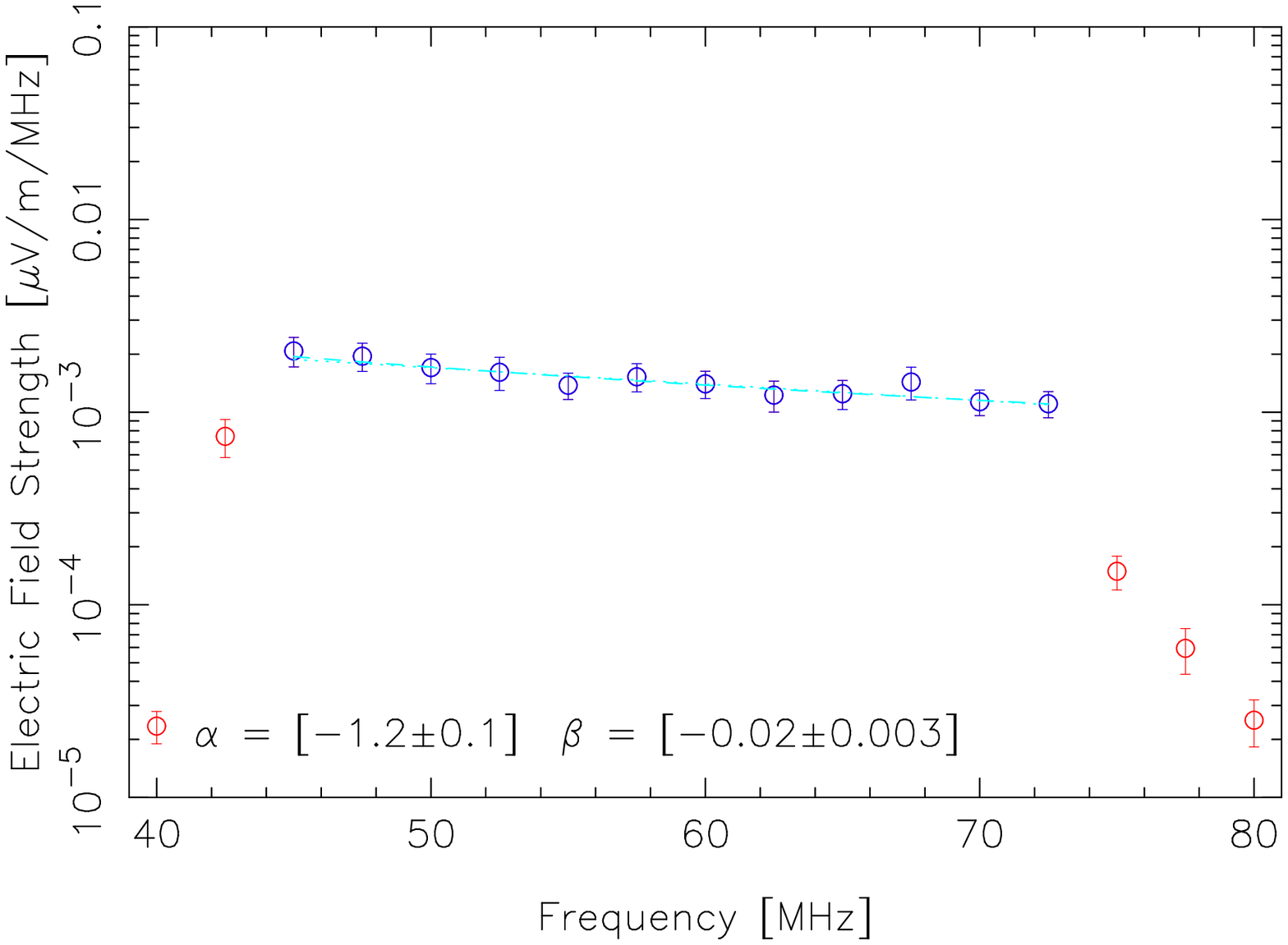} 
\par\end{centering}

\caption{\label{fig:cr_noiseFspec}System noise spectrum of the LOPES beam
to the zenith in units of electric field strength calculated on 664
events of 1~ms recorded over a period of one year (11/2005 - 11/2006).
The numbers on the bottom of the plot are, respectively, the slope
parameters $\alpha$ and $\beta$ for a power-law function and an
exponential function fitted to the spectrum (dashed line).}

\begin{centering}
\vspace{20pt}
\par\end{centering}

\begin{centering}
\includegraphics[width=0.8\columnwidth]{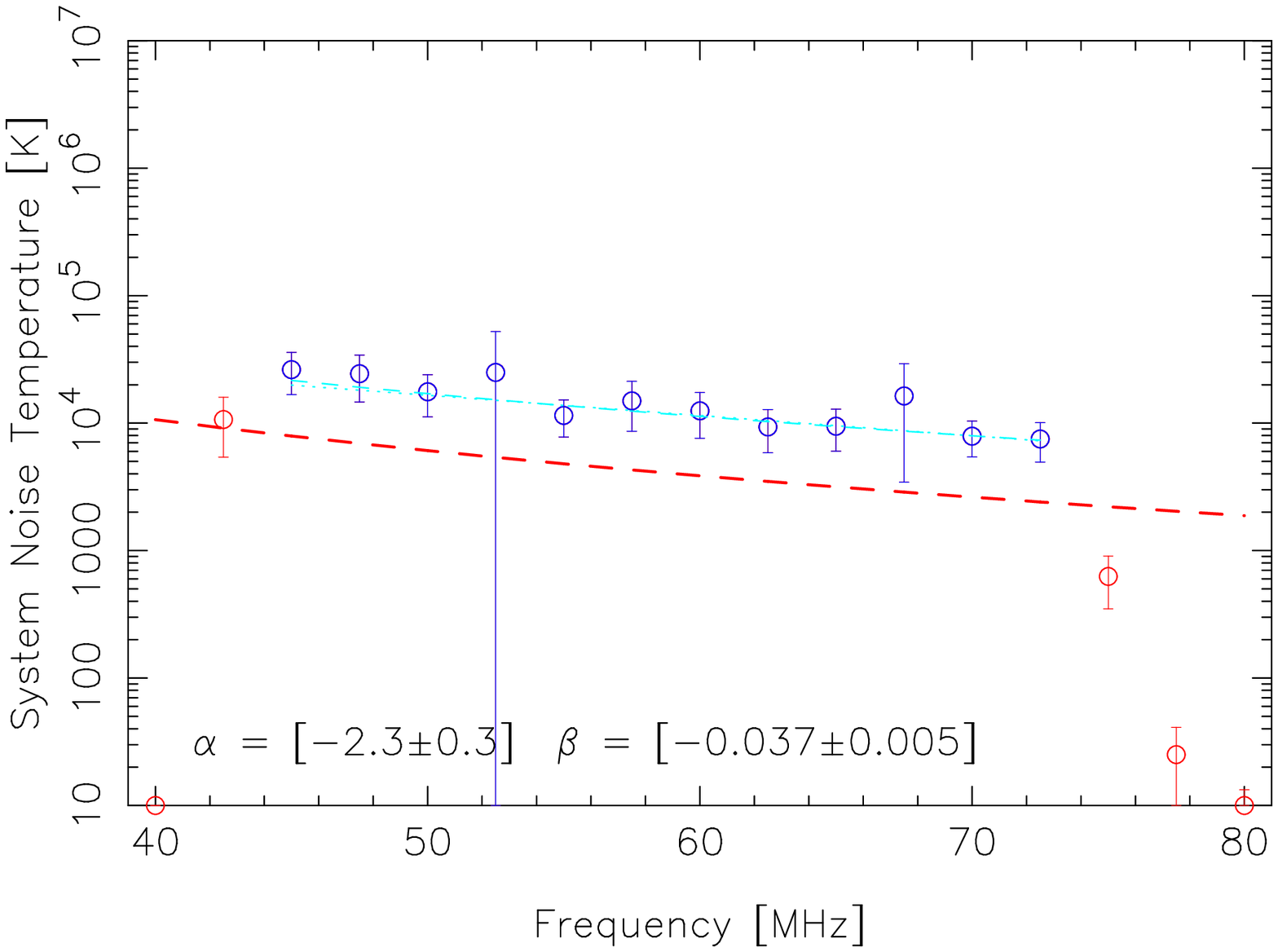}
\par\end{centering}

\caption{\label{fig:cr_noiseTspec}Spectrum of the LOPES beam to the zenith
in units of noise temperature calculated on 664 events of 1~ms recorded
over a period of one year. The numbers on the bottom of the plot are
the slope parameters $\alpha$ and $\beta$ for a power-law function
and an exponential function fitted to the spectrum (dashed line).
The galactic noise is plotted for comparison (lower dashed line).}

\end{figure}

\begin{figure}
\begin{centering}
\includegraphics[width=1\columnwidth]{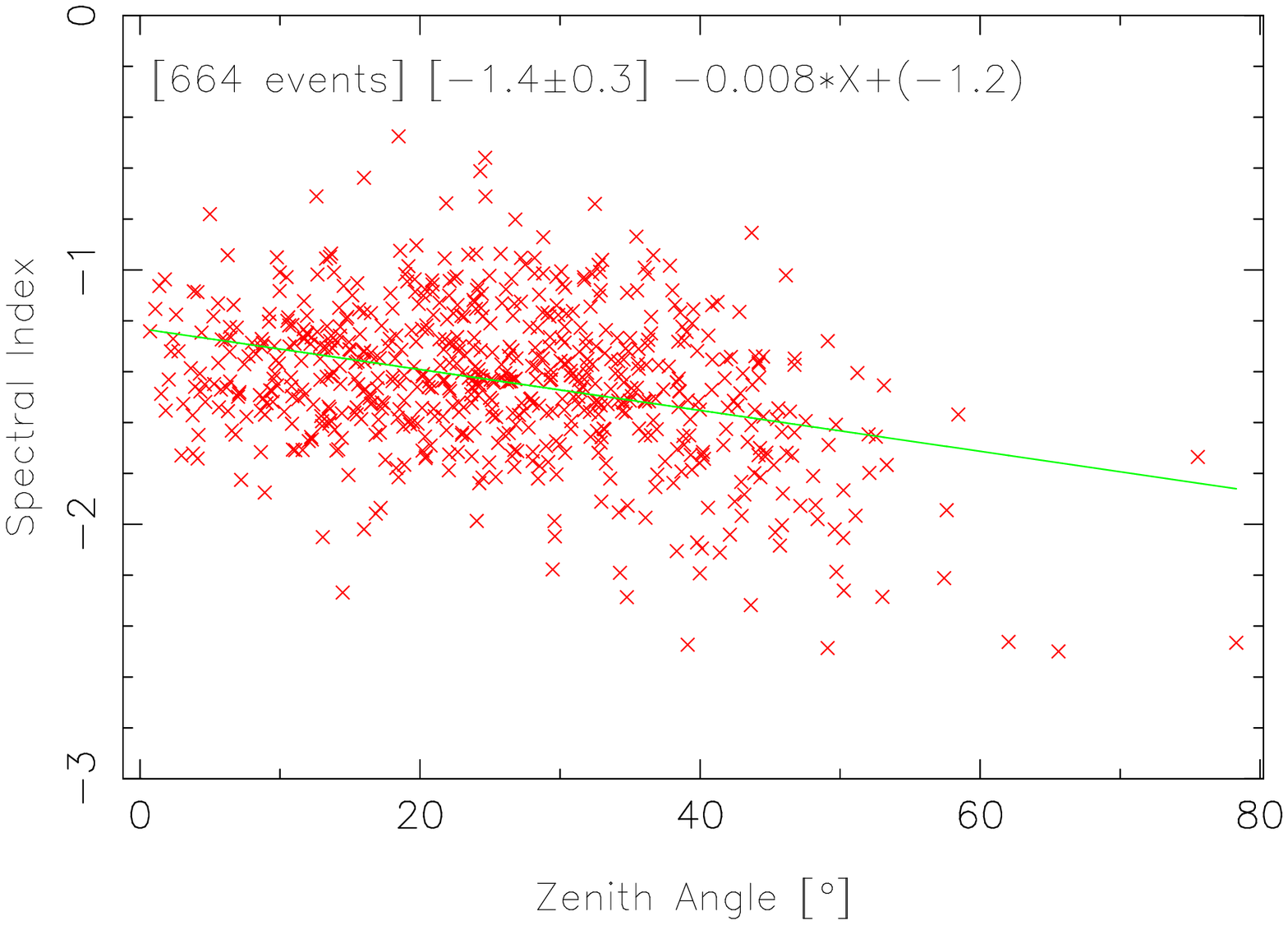} 
\par\end{centering}

\caption{\label{fig:crspec_NindVSze}Spectral index of 664 noise spectra as
a function of zenith angle. The average slope is $\alpha=-1.4\pm0.3$
and the fitted line is described by $-0.008\times\left(90-\theta/\unit{^{\circ}}\right)-1.2$.}

\end{figure}

Here, $\hat{\epsilon}_{\nu}$ is the electric field amplitude for
the frequency bin $\nu$, $\alpha_{f}$ is the slope parameter for
the measured spectrum obtained from a fit to the slopes in Fig. \ref{fig:crspec_NindVSze}
of $-0.008\times\left(90-\theta/\unit{^{\circ}}\right)-1.2$, and
$\alpha_{r}$ is the slope of the reference spectrum in the zenith
(Fig. \ref{fig:cr_noiseFspec}) of $-1.2$.

\subsection{Pulse amplitude\label{sub:cr_amplitude}}

\begin{figure}
\begin{centering}
\includegraphics[width=1\columnwidth]{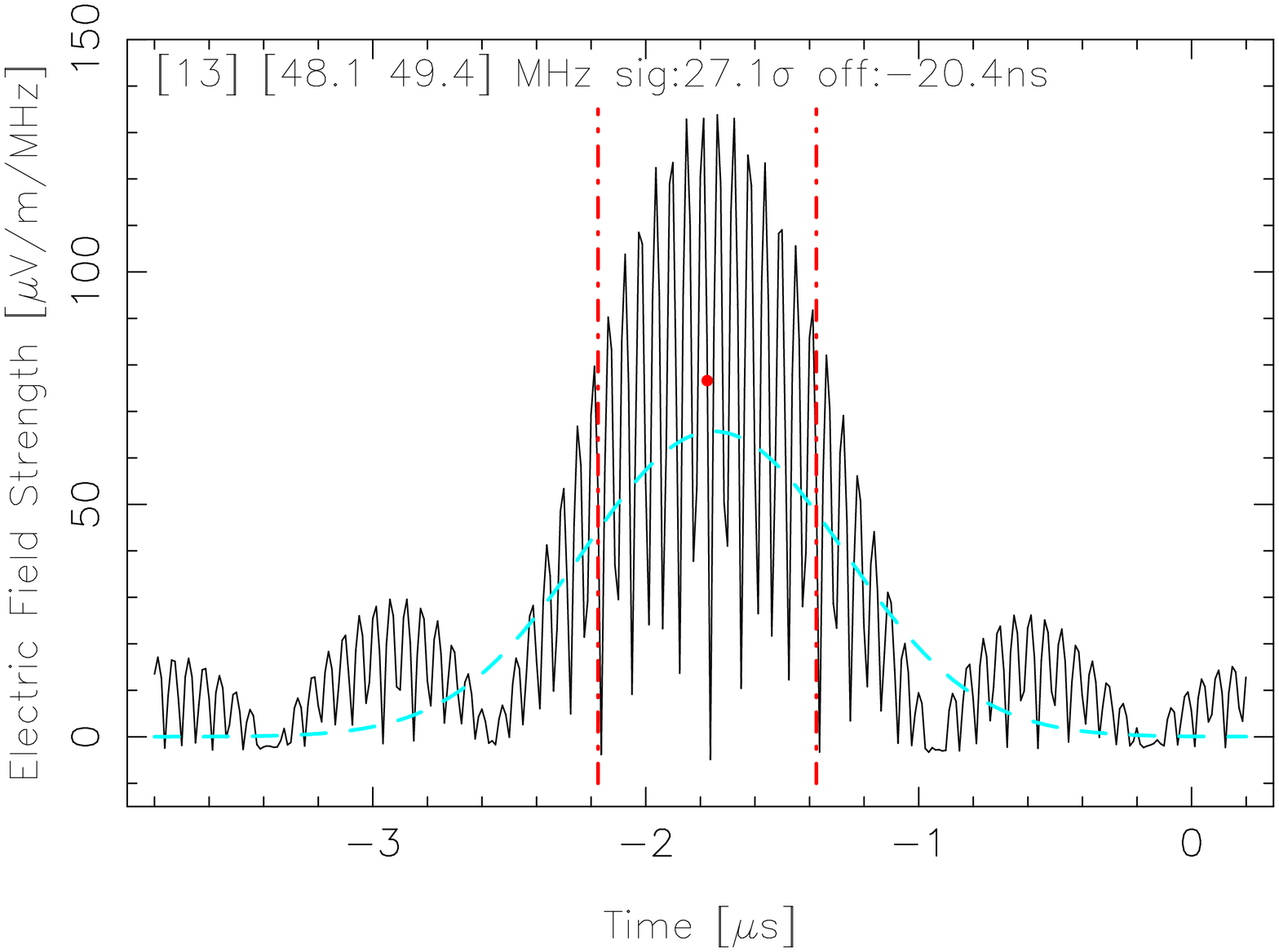} 
\par\end{centering}

\caption{\label{fig:cr_ccbeam}Cross-correlation beam filtered in a frequency
sub-band (thin solid line) and scaled full-band Gauss-fit (dashed).
The vertical lines (dash-dotted) indicate the beginning and end of
the part of the cc-beam that was averaged and the dot  in the middle
indicates the resulting value of $76.6~\unit{µV~m^{-1}MHz^{-1}}$.}

\end{figure}

The field strength amplitude of the frequency filtered cc-beam was
determined by averaging it over a time window with the size of the
inverse sub-band. As an example, a broadened pulse for a sub-band
of $\sim1.3~\unit{MHz}$ is plotted in Fig. \ref{fig:cr_ccbeam}.
The values in the top left corner of the plot read: event number 13,
frequency sub-band from $48.1~\unit{MHz}$ to $49.4~\unit{MHz}$;
E-field peak SNR of $27.1\sigma$ of the frequency bin and the offset
between the full-band peak maximum and the sub-band peak maximum of
$-20.4~\unit{ns}$. The dot lies at a higher field strength, since
its frequency bin lies below the center frequency of the full-band
beam of $60~\unit{MHz}$. The cc-beam shows correlation fringes for
the center frequency of the sub-band. For comparison, the Gauss-fit
on the smoothed full-band peak was broadened in width by the total
number of frequency bins of $24$. The plotted pulse for a sub-band
below the center frequency of the LOPES band is higher than the full-band
pulse, which is consistent with the negative slope in the resulting
spectrum.

\subsection{Uncertainty estimation}

The uncertainty of the determined electric field strength is affected
by the direction and frequency dependent gain factors, phase uncertainties
introduced by the electronics and the background noise:\begin{eqnarray}
\Delta\epsilon & = & \sqrt{\left(\frac{\delta\epsilon}{\delta t}\right)^{2}+\left(\frac{\delta\epsilon}{\delta\alpha}\right)^{2}+\left(\frac{\delta\epsilon}{\delta\phi}\right)^{2}}.\label{eq:cr_tot_stat_err}\end{eqnarray}

Here, $\left(\frac{\delta\epsilon}{\delta t}\right)$ is the statistical
uncertainty (RMS) of the background noise (see Sect. \ref{sub:cr_background_noise});
$\left(\frac{\delta\epsilon}{\delta\alpha}\right)$ is the statistical
uncertainty of the E-field dependence on the beam-forming direction;
and $\left(\frac{\delta\epsilon}{\delta\phi}\right)$ is an estimated
uncertainty of 5\% introduced by phase errors remaining after delay
calibration (\citealt{horneffer06}).

The dependence of the E-field on the beam-forming direction is implemented
in the gain calibration and a statistical uncertainty is estimated
for each direction and frequency sub-band calculating a gradient $\frac{\delta G}{\delta\alpha}$
on the beam-pattern:\begin{eqnarray}
\epsilon & = & k\cdot\frac{1}{\sqrt{G\left(\alpha\right)}}\nonumber \\
\left(\frac{\delta\epsilon}{\delta\alpha}\right) & = & k\cdot-\frac{1}{2}G\left(\alpha\right)^{-1.5}\cdot\frac{\delta G}{\delta\alpha}\label{eq:cr_gain_err}\\
 & = & k\cdot\frac{1}{\sqrt{G\left(\alpha\right)}}\cdot-\frac{1}{2G\left(\alpha\right)}\cdot\frac{\delta G}{\delta\alpha}.\nonumber \end{eqnarray}

Here, $\epsilon$ is the electric field strength, $k$ is a constant
factor (see Eq: \ref{eq:cr_gain_cal}), $G\left(\alpha\right)$ is
the direction dependent gain factor, and $\frac{\delta G}{\delta\alpha}$
is its gradient.

The determined uncertainties are consistent within the uncertainties
applicable to earlier LOPES results (\citealt{horneffer06}). For
all 23 analyzed events, the average uncertainties are listed in Table
\ref{tab:cr_errs}. 

\begin{table}
\begin{centering}
\caption{\label{tab:cr_errs} List of uncertainties on the determined electric
field strengths.\vspace{5pt}}

\par\end{centering}

\begin{centering}
\begin{tabular}{|c|c|c|c|}
\hline 
Method & RMS  & Gain Uncert. & Phase Uncert.\tabularnewline
\hline
\hline 
cc-beam & 1.59 (7.1\%) & 0.35 (1.56\%) & 1.13 (5\%)\tabularnewline
\hline 
f-beam & 0.78 (3.8\%) & 0.32 (1.56\%) & 1.01 (5\%)\tabularnewline
\hline
\end{tabular}
\par\end{centering}

\vspace{5pt} These values are calculated from all 23 analyzed events
in units of $\unit{µV~m^{-1}MHz^{-1}}$. The values in brackets are
the averaged percentages of the actual electric field values.
\end{table}

\section{Results\label{sec:cr_fieldspec}}

For the electric field spectra, a frequency resolution of 2.5~MHz
(16~bins over the 40~MHz band) was chosen as a compromise between
pulse broadening and sufficient spectral resolution to reveal phase
uncertainties. The phase uncertainties are mainly caused by narrowband
RFI and deviations from the frequency dependent gain calibration.
The radio spectra from Monte Carlo simulations were parametrized with
an exponential (\citealt{huege05app}). For comparison, an exponential
$\epsilon_{\nu}=K\cdot\mbox{exp}(\nu/\unit{MHz}/\beta)$ and a power-law
form $\epsilon_{\nu}=K\cdot\nu^{\alpha}$ were fitted to the data
of the single and the average spectra obtained by both methods and
corrected as described in Sect. \ref{sub:cr_background_noise}. However,
the $\chi_{red}^{2}$ of both fit types turns out to be the same and
thus both functions fit equally well. The resulting fit values can
be found in Sect. \ref{sub:cr_avgspec} and Table \ref{tab:cr_spectra_par}
at the end of the paper.

\subsection{Single event spectra\label{sub:crspec_spec}}

Radio spectra were calculated for 23 LOPES events. Characteristic
values for each event are listed in Table \ref{tab:cr_spectra_par},
including the event number; the event date; the electric field strength
measured in the whole LOPES band from the spectra $\epsilon$; the
spectral index $\alpha$; the exponential fit parameter $\beta$;
the width of the Gauss-fitted full-band pulse $\Delta t$; the E-field
peak SNR from the spectra in sigma $\sigma$; the estimated primary
energy $E_{p}$ (\url{http://www-ik.fzk.de/~ralph/CREAM1.php}, \citealt{glasstetter05icrc},
only reliable down to $48\unit{^{\circ}}$ elevation, no values are
given below $20\unit{^{\circ}}$ elevation); the direction (azimuth
$\phi$ and elevation $\theta$); the angle of the shower axis with
the geomagnetic field $\vartheta$; the average distance of the antennae
to the shower core $d$; and the number of antennae used for beam-forming
$N$. The values for $\epsilon$, $\alpha$, $\beta$, $\Delta t$,
and $\sigma$ are given as the average result from the cc-beam and
the f-beam. The error was chosen as the maximum from both methods.

Additionally, six single event spectra for both methods are plotted
in Fig. \ref{fig:cr_efieldspecs}. For each spectrum an average noise
spectrum was calculated by FFT of 50 blocks offset from the cosmic-ray
radio pulse, which is plotted as well. Furthermore, two spectra from
Monte Carlo simulations are added to each plot for an inclined shower
with $\theta=45\unit{^{\circ}}$ and for a vertical shower with $\theta=90\unit{^{\circ}}$
(\citealt{huege07}). The showers were simulated for a proton with
a primary energy of $E_{p}=10^{17}~\unit{eV}$, an azimuthal viewing
angle of $\phi=45\unit{^{\circ}}$ from the observer measured at a
shower core distance of 100~m and were scaled to have the same average
as the single spectra for comparison. 

The values on the top of the plots in Fig. \ref{fig:cr_efieldspecs}
read; in the 1st line: the event number, the measured electric field
strength from the spectra, the E-field peak SNR from the spectra in
sigma, the estimated primary energy (\citealt{glasstetter05icrc}),
the width of the Gauss-fitted full-band pulse, the direction (azimuth
angle and elevation angle) and the geomagnetic field angle; in the
2nd line: the fit parameters $\alpha$ and $\beta$ for the spectral
slope. All the E-field values, slope values, sigma values, and width
values in the table and the plots in the appendix are given as the
average of the f-beam and the cc-beam method together with the maximum
uncertainty.

The six plotted events have the following special characteristics.
The analyzed events indicate that with larger pulse width, lower frequencies
dominate and steepen the spectrum (see event {[}16] with a width of
$51~\unit{ns}$ and a flat spectrum, versus event {[}7] with a width
of $101~\unit{ns}$ and a steep spectrum). The latter event was found
to be recorded during thunderstorm activity in the vicinity of LOPES.
Strong electric fields in the clouds may have enhanced the radio emission
and broadened the pulse (\citealt{buitink07}). However, the sample
is too small and no significant dependence was found. Furthermore,
the measured electric field amplitude increases with increasing angle
of the shower axis with the geomagnetic field, as expected, according
to the geomagnetic emission mechanism (compare event {[}10] with $11~\unit{\mu V~m^{-1}MHz^{-1}}$
and $51\unit{^{\circ}}$, versus event {[}5] with $20~\unit{\mu V~m^{-1}MHz^{-1}}$
and $89\unit{^{\circ}}$). In addition, two events with the largest
electric field amplitude are plotted in Fig. \ref{fig:cr_efieldspecs},
which have the smallest uncertainties (see event {[}20] and {[}21]
with geomagnetic angles of $78\unit{^{\circ}}$ and $70\unit{^{\circ}}$).

\subsection{Average field spectrum\label{sub:cr_avgspec}}

\begin{figure}
\begin{centering}
\includegraphics[width=1\columnwidth]{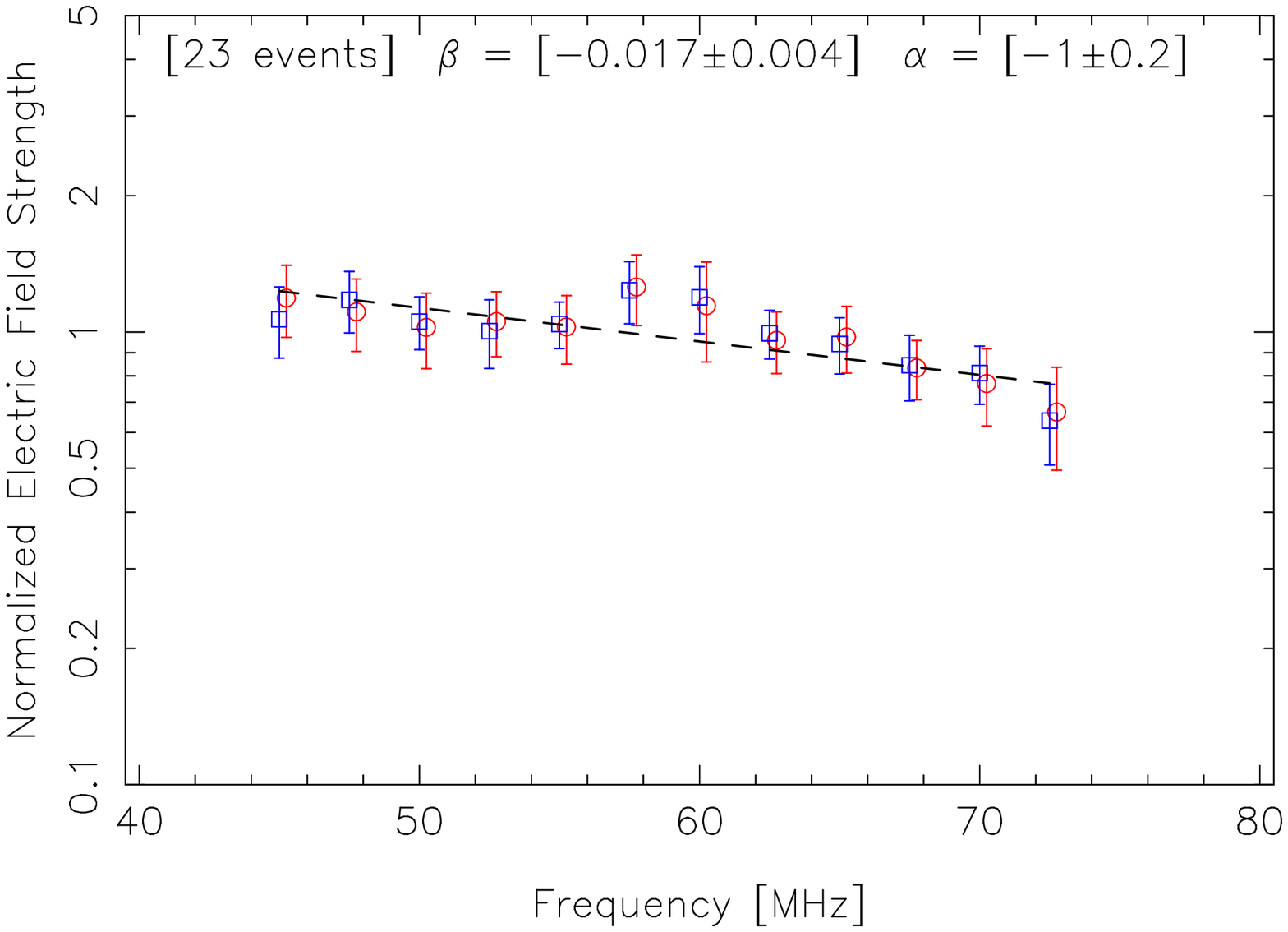}
\par\end{centering}

\caption{\label{fig:crspec_avgspec} Comparison of average cosmic-ray electric
field spectra obtained with 23 LOPES events by two different methods.
Before averaging, the single spectra are normalized to unity. The
frequency bin values are determined on the cc-beam (circles) and the
Fourier transformation of the f-beam (squares) are fitted with one
exponential, visible as a straight line in this log-linear plot (dashed
line).}

\end{figure}

An average field spectrum was calculated to obtain an average spectral
slope for the 23 selected LOPES events (Fig. \ref{fig:crspec_avgspec}).
For this plot, the single event spectra were normalized by their mean
to unity to make them energy independent and then they were averaged.
The numbers on the top indicate the number of events taken into account
for the average spectrum and the parameters from the spectral fits
to the plotted spectra. The average exponential fit with the parameter
$\beta$ is plotted for the two methods.

The exponential parameter is obtained with $\beta=\mathbf{\mathnormal{-0.017}}\pm0.004$
and the spectral index is obtained with $\alpha=-1.0\pm\mathbf{\mathnormal{0.2}}$.
These fit parameters apply to an average zenith angle of the 23 events
of $53\unit{^{\circ}}$ and an average distance of the antennae from
the shower core position of 76~m.

The average spectrum measured confirms basic expectations of \citet{falcke03},
but it is not consistent within the uncertainties and it is slightly
steeper than the slope obtained from Monte Carlo simulations based
on air showers simulated with CORSIKA (Cosmic Ray Simulations for
KASCADE). The fit parameters from simulations of a $45~\unit{^{\circ}}$
inclined air shower are $\beta=-0.0085$ and $\alpha=-0.49$; and
the values for a vertical shower are $\beta=-0.012$ and $\alpha=-0.7$
(\citealt{huege07}). For a closer comparison, simulations on single
event basis are necessary taking into account polarization characteristics
of the LOPES antennae and those simulations have not yet been performed.

For further comparison, the spectral index recently measured by the
CODALEMA group for a single event spectrum is $\alpha=-1.5\pm0.2$
(\citealt{codalema06}), which is consistent with the average slope
we obtained for the uncorrected spectra.

\subsection{Discussion of methods}

The major difference in the determination of the electric field strength,
between the calculation of the spectra and the standard analysis software,
is that the latter fits a Gaussian to the pulse. The standard software
assumes the pulse to be Gaussian and any deviation of the real cosmic-ray
radio pulse shape will cause deviations in the resulting electric
field strength. Thus, the method used here for the spectra is more
sensitive to the shape of the radio pulses. The electric field strengths
determined from the spectra agree within \textpm{}30\% with those
obtained with the standard analysis (one sigma statistical uncertainty).

The two methods for the spectra determination are in statistical agreement.
The noise on the cc-beam is a few percent ($\sim$3 \%) larger than
the noise on the f-beam (see Sect. \ref{sub:cr_background_noise}).
Furthermore, the f-beam method is more accurate, since the statistical
uncertainties of all events averaged by weighting is only half of
the value obtained for the cc-beam method. Additionally, the $\chi^{2}$
of the fits on the spectra from the f-beam method are on average an
order of magnitude better than for the cc-beam method.

\subsection{Pulse width analysis}

\begin{figure}
\begin{centering}
\includegraphics[width=1\columnwidth]{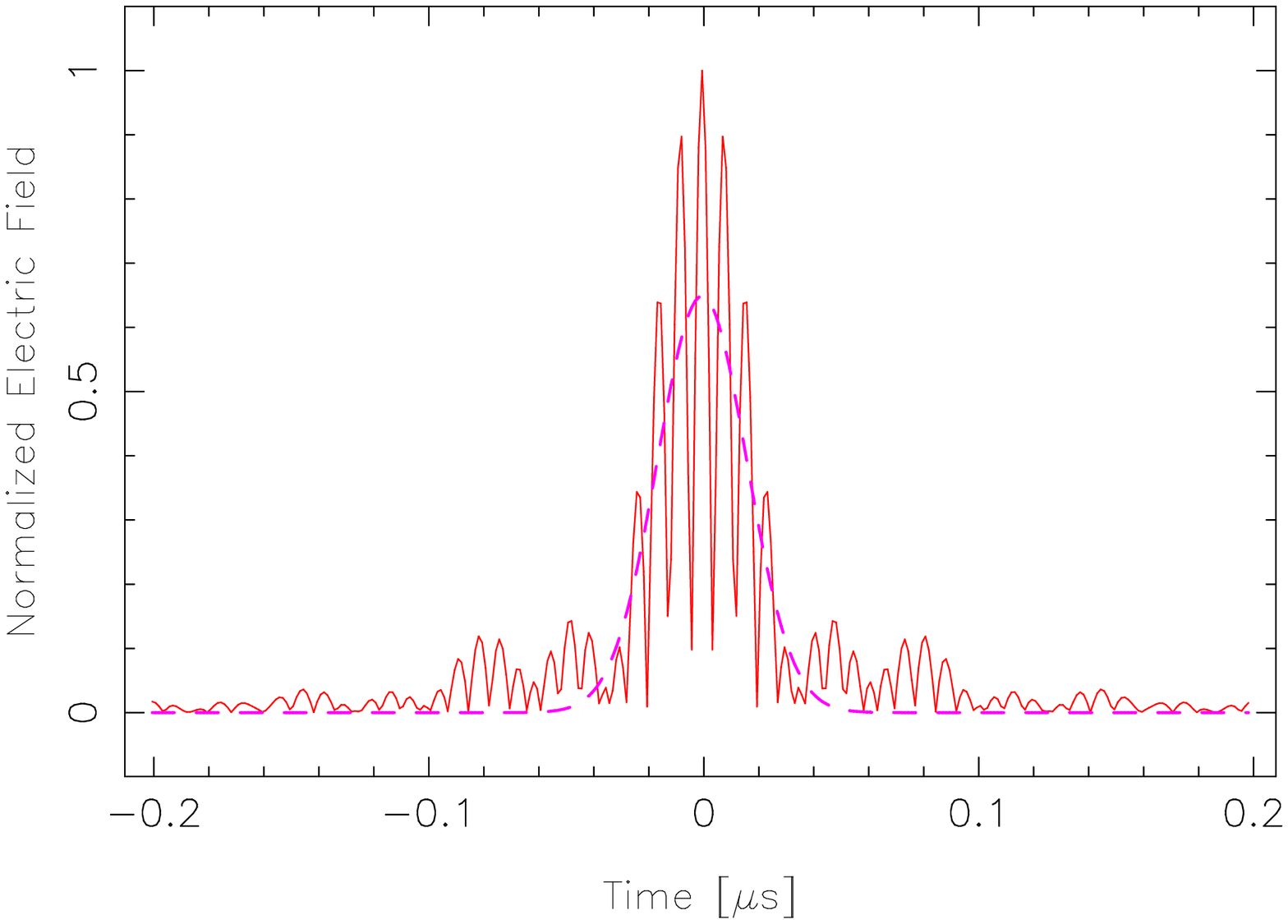} 
\par\end{centering}

\caption{\label{fig:cr_min_pulse_width}Pulse obtained by inverse Fourier transformation
of the amplitudes of the average electric field spectrum (solid line).
The width of the Gaussian fit results in $40~\unit{ns}$ (dashed line).
The amplitude is normalized to unity.}

\end{figure}

\begin{figure*}
\caption{\label{fig:cr_efieldspecs}Single event cosmic-ray electric field
spectra determined on six LOPES example events. Each plot includes
a spectrum determined on the cc-beam (circles) for 12 sub-bands, a
spectrum determined on the f-beam (squares) by FFT of 32 time-samples
around the radio pulse, and a noise spectrum determined by FFT on
50 blocks of 32 samples offset from the pulse (squares in the middle
of each plot). The spectra of the cosmic-ray air shower radio pulse
are fitted with an exponential, visible as a straight line in these
log-linear plots (cc-beam: dash-dotted \& f-beam: dash-dot-dot). In
addition, the simulated data of an inclined shower (dotted) and a
vertical shower (dashed) are plotted. For more details see the second
to last paragraph of Sect. \ref{sub:crspec_spec}.\vspace{10pt}}

\begin{centering}
\includegraphics[width=0.5\linewidth]{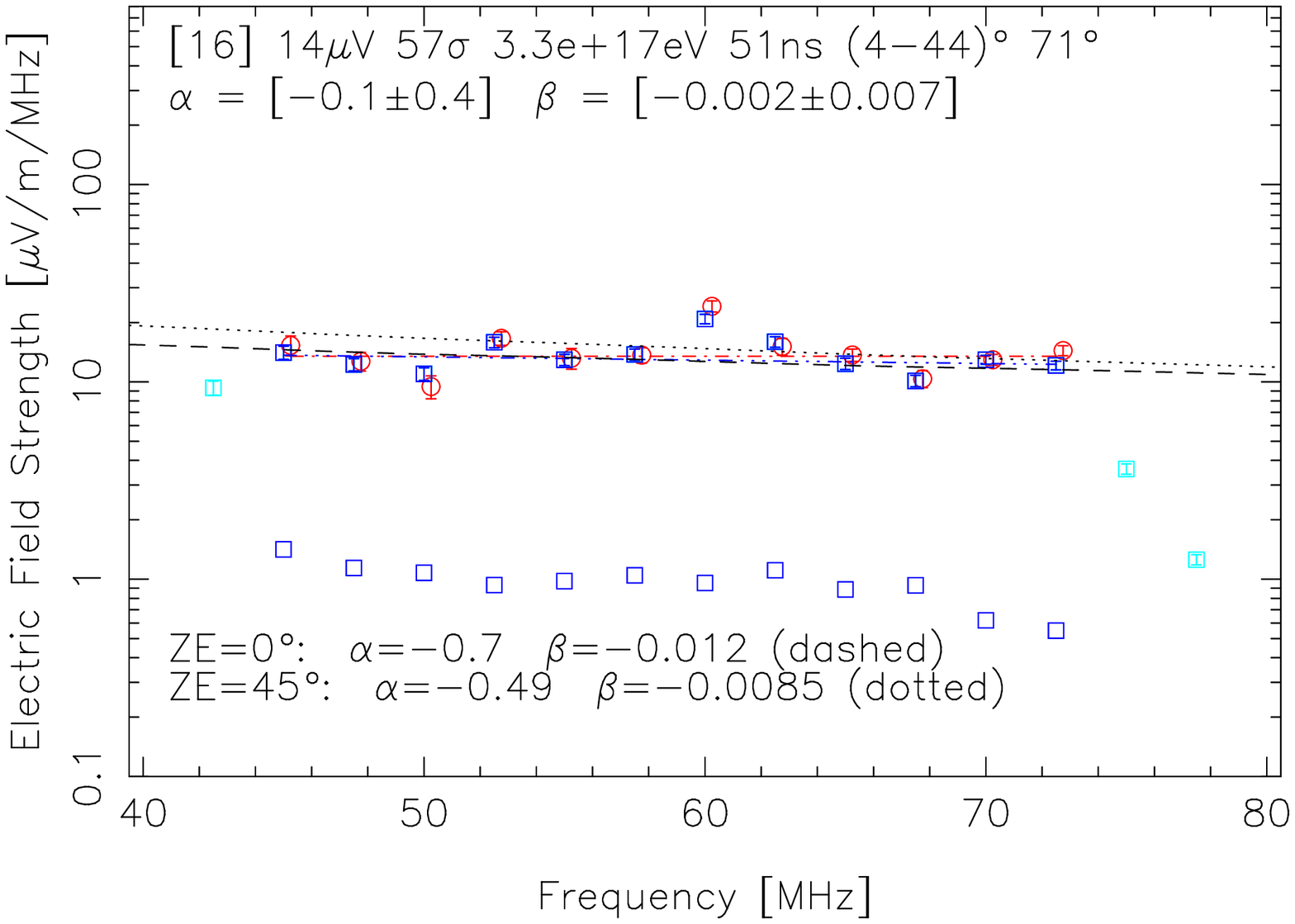}\includegraphics[width=0.5\linewidth]{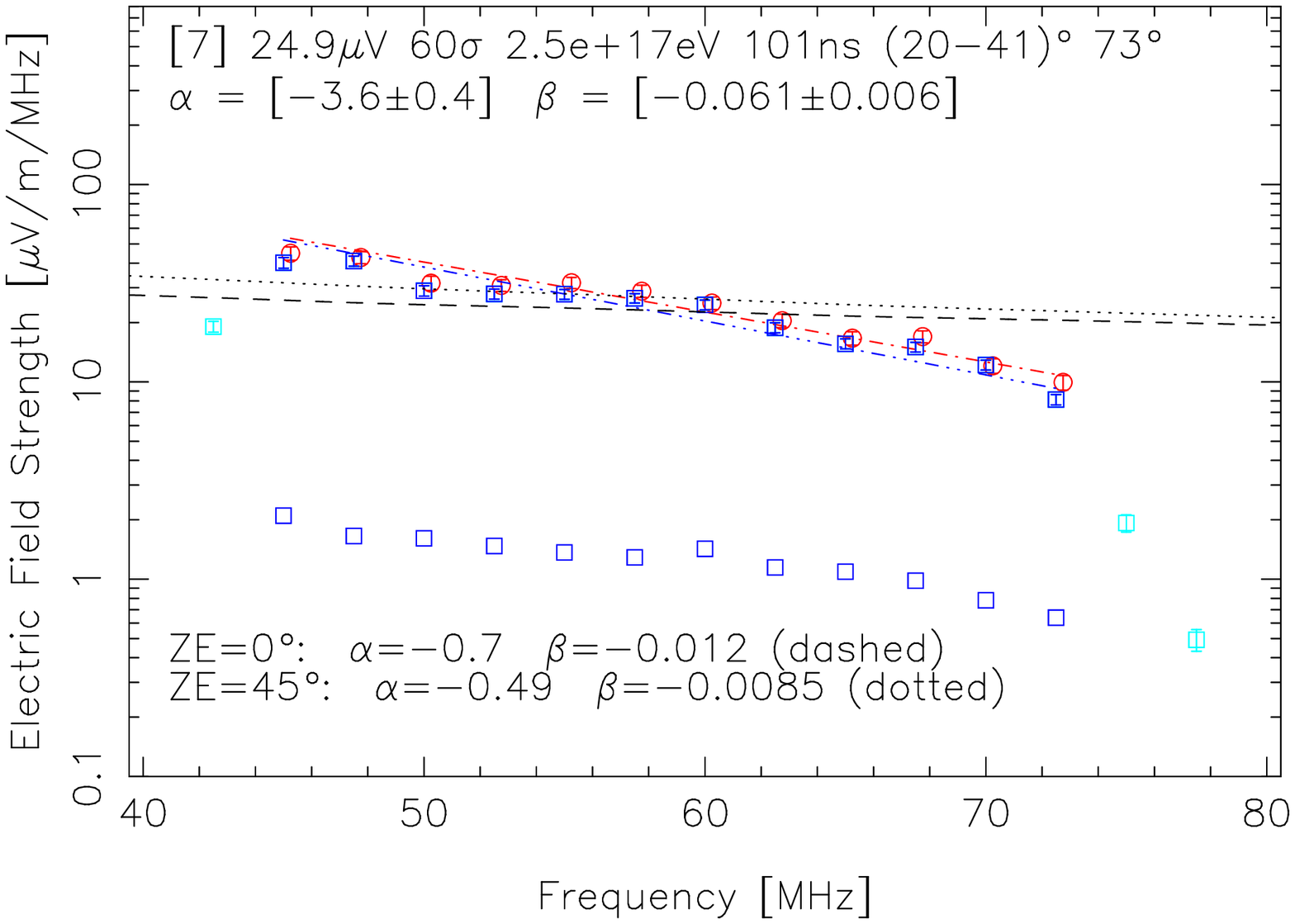}
\par\end{centering}

\begin{centering}
\includegraphics[width=0.5\linewidth]{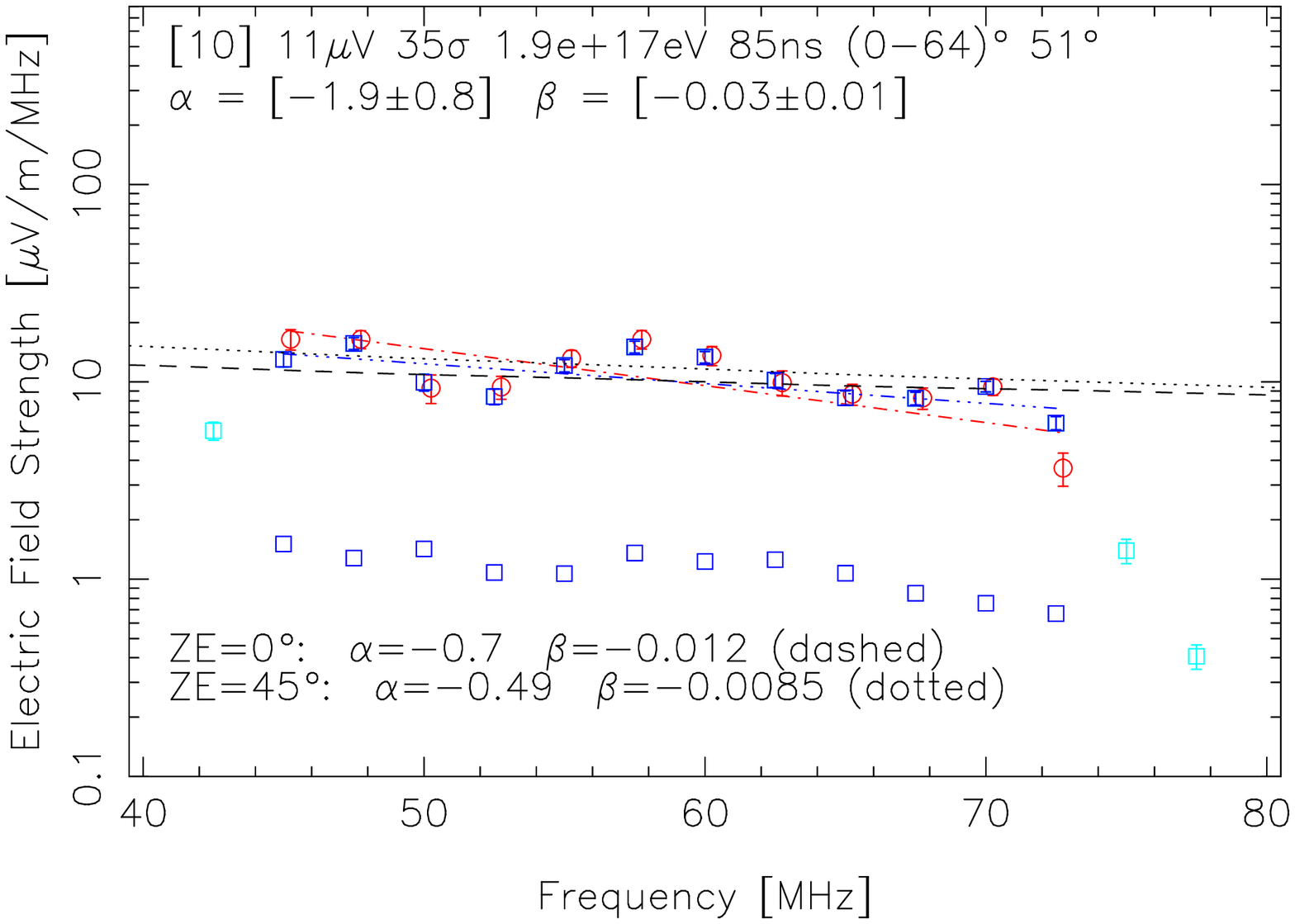}\includegraphics[width=0.5\linewidth]{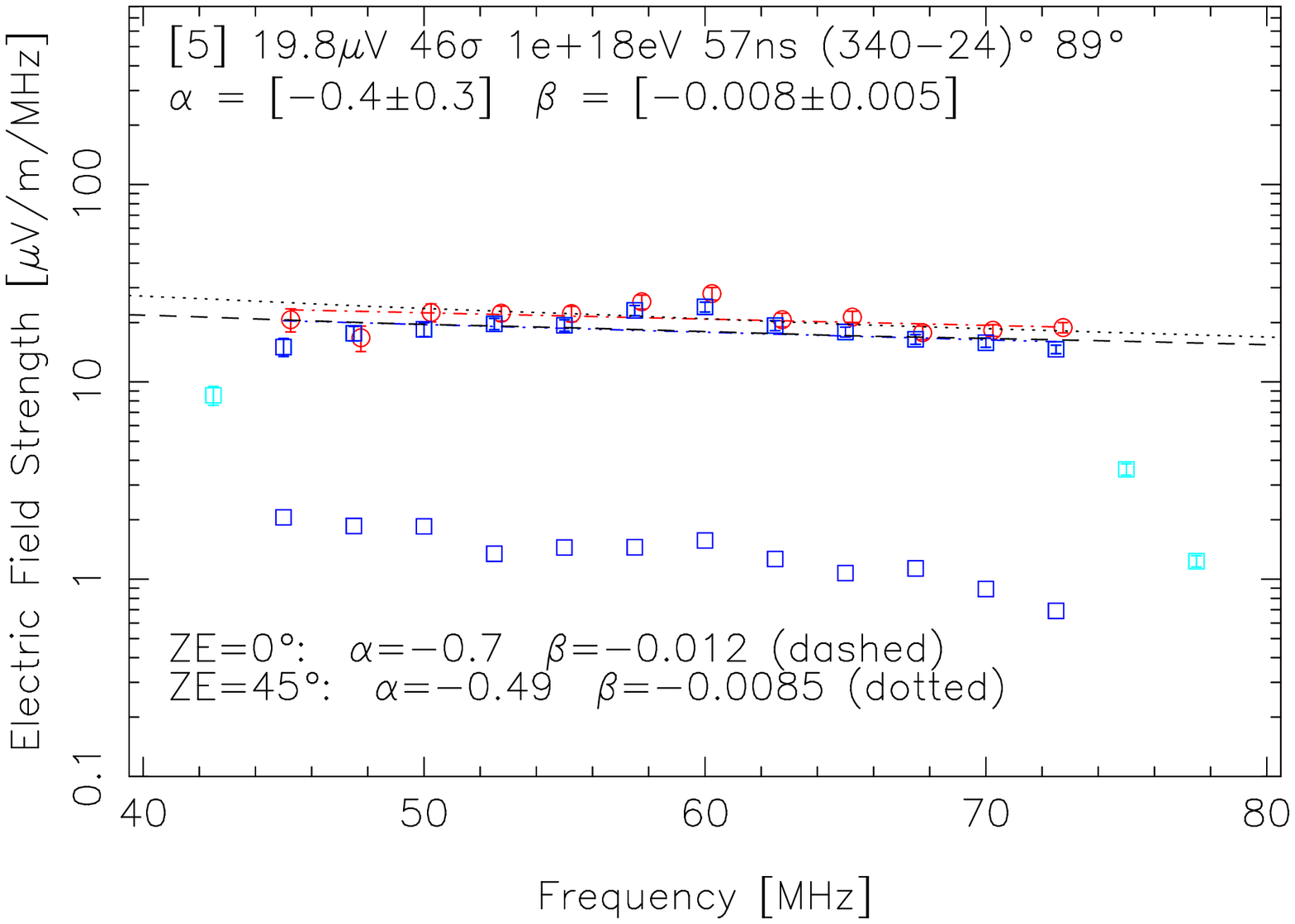}
\par\end{centering}

\begin{centering}
\includegraphics[width=0.5\linewidth]{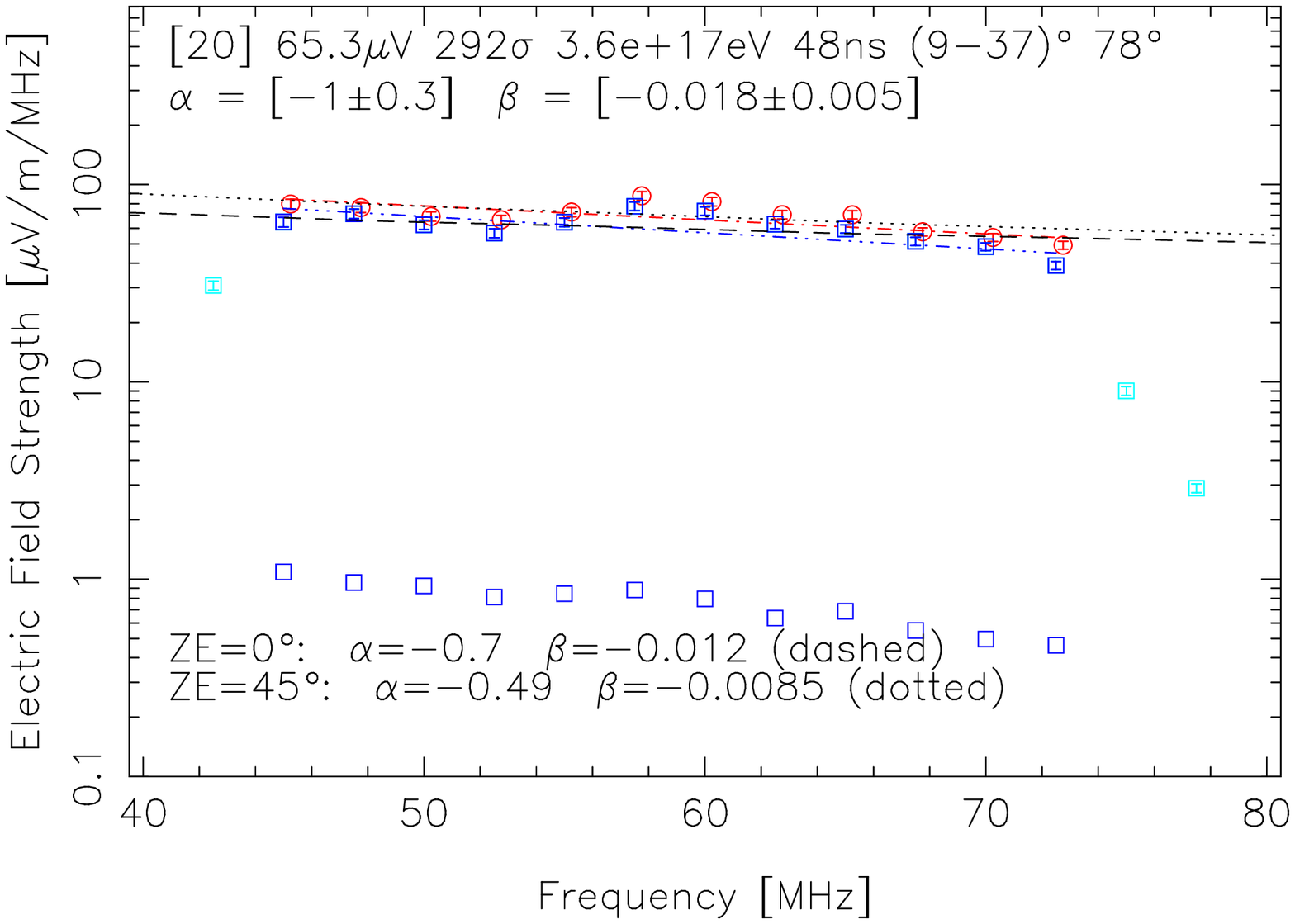}\includegraphics[width=0.5\linewidth]{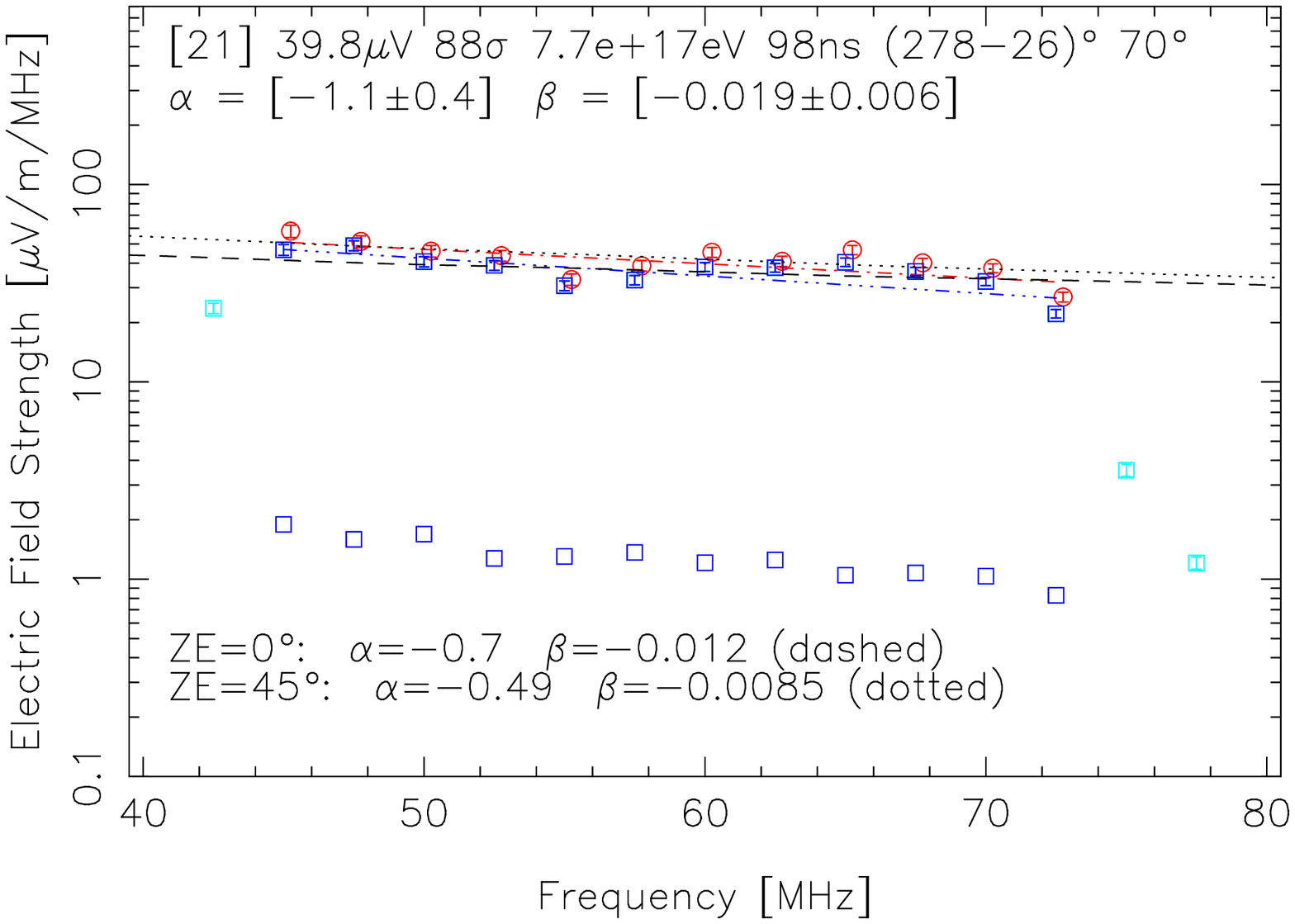}
\par\end{centering}
\end{figure*}

The resulting average radio spectrum can be used to obtain an average
shape for the measured radio pulses by inverse Fourier transformation.
However, the phases are modified by the electronics, which at the
moment, we cannot correct for. Thus, a lower limit on the pulse width
can be provided by applying an inverse Fourier transformation on the
measured spectral amplitudes (see Fig. \ref{fig:cr_min_pulse_width}).
The obtained pulse was upsampled in frequency to increase the number
of samples for a fit. The amplitude was normalized to unity, since
it is not comparable with earlier E-field values. A Gaussian fit on
the obtained pulse results in a full-width-half-maximum (FWHM) of
$40~\unit{ns}$. As expected, this minimum is slightly smaller than
the impulse response of the electronics of $\sim57~\unit{ns}$. The
average pulse width of the analyzed events resulted in $60\pm20~\unit{ns}$
(one sigma statistical uncertainty). This value was calculated on
41 events, for which the difference between the f-beam width and the
cc-beam width was not larger than two samples ($25~\unit{ns}$). The
two samples are a quality criterion. The simulated and measured pulse
width set an upper limit of $60~\unit{ns}$ and the minimum pulse
width a lower limit of $40~\unit{ns}$ for the single selected LOPES
pulses. Therefore, the obtained spectral slopes provide an upper limit,
possibly broadened by unknown phase uncertainties, which were estimated
for the uncertainties in the spectra.

\section{Conclusion and outlook}

With a sample of 23 strong LOPES events, we measured the radio spectrum
received from cosmic-ray air showers detected with up to 24 simple
dipole antennae oriented in the East-West direction over a frequency
band of 40~MHz. The spectra show that a direct Fourier transformation
can be performed on the beam-formed radio pulses measured with LOFAR.
The accuracy with which the spectral amplitudes can be obtained is
limited by the instrument noise and phase uncertainties. Furthermore,
the quality of the spectral slope is limited by the quality of the
antenna gain model, which was simulated and measured in several calibration
campaigns.

The average slope of the spectra obtained with LOPES confirms basic
expectations, but it is not consistent within uncertainties and it
is slightly steeper than the slope obtained from Monte Carlo simulations
based on air showers simulated with CORSIKA. The simulations show
for the LOPES band flatter spectra to higher zenith angles and larger
distances from the shower core position, and they show steeper spectra
for increasing the azimuth from direction North to direction East.

As expected, the spectral slopes of the selected sample of events
depend on the length of the pulse, where longer pulses result in steeper
spectra. However, taking into account the low number statistics, the
spectra do not show a significant dependence of the slope on the electric
field amplitude, the azimuth angle, the zenith angle, the curvature
radius, nor on the average distance of the antennae from the shower
core position.

According to the obtained spectral slopes, the maximum power is emitted
below 40~MHz. Furthermore, the decrease in power to higher frequencies
indicates a loss in coherence determined by the shower disc geometry
and the longitudinal distribution of particles therein. In Monte Carlo
simulations, the latter was found to follow a broadened gamma-distribution
having a long tail to several meters in shower thickness, which confirms
a smooth loss of coherence to higher frequencies (see Fig. 7 in \citealt{huege07}).

The selected sample of events contained one event that occurred during
thunderstorm activity in the vicinity of LOPES. This thunderstorm
event had a large pulse width and a very steep spectrum, suggesting
that LOPES measured a geoelectric effect for the first time.

For the study of the lateral dependence of the air shower radio spectrum,
as a function of distance to the shower axis, a larger array with
longer baselines and more sensitive antennae is needed. LOFAR will
have a very dense core with baselines up to 500 meters, a collecting
area of $\sim10000~\unit{m^{2}}$ at 75~MHz, and a frequency range
from 10~MHz to 290~MHz sampled at 200~MHz, and thus will be perfectly
suited to improve on results obtained here. Most importantly, we will
probe the coherence of the air shower radio signal to higher frequencies
to infer on the geometry of the shower disc. Understanding the shower
geometry will allow us to determine the primary particle species using
radio only.

Furthermore, within the Pierre Auger collaboration, a sparse radio
antenna array is under development (\citealt{berg07}). The antennae
will be tested in the field of the Pierre Auger Observatory (PAO,
\url{www.auger.org}). The PAO measures cosmic-ray air showers with
particle detectors in an area of 3000 square kilometers with distances
between the particle detector stations of 1.5~km. Therefore, the
PAO responds to a higher primary particle energy range than LOPES,
which lies beyond $10^{18}~\unit{eV}$.

A next step will be the simulation of complete air showers based on
parameters provided by KASCADE and LOPES, including the hardware response
of LOPES. Such simulations will allow a close comparison of theory
and experiment.\linebreak

\emph{\small Acknowledgments.}{\small{} Andreas Nigl gratefully acknowledges
a grant from ASTRON, which made this work possible. LOPES was supported
by the German Federal Ministry of Education and Research. The KASCADE-Grande
experiment is supported by the German Federal Ministry of Education
and Research, the MIUR and INAF of Italy, the Polish Ministry of Science
and Higher Education and the Romanian Ministry of Education and Research.}{\small \par}

\bibliographystyle{aa}
\bibliography{anigl}

\clearpage \onecolumn

\markboth{}{}

%
\begin{sidewaystable}
\begin{raggedright}
\caption{\label{tab:cr_spectra_par} Event parameter list for all 23 analyzed
events. For more details see first paragraph of Sect. \ref{sub:crspec_spec}.}

\par\end{raggedright}

\begin{raggedright}
\begin{tabular}{|c|l|c|c|c|c|c|c|c|c|c|c|}
\hline 
\# & \multicolumn{1}{l|}{Date (Filename)} & E-field & Index & Expo. & Width & Sig. & Energy & AZEL & Geom. Ang. & Dist. & Ant.\tabularnewline
\hline 
 & Symbol & $\epsilon$ & $\alpha$ & $\beta$ & $\Delta t$ & $\sigma$ & $E_{p}$ & $\phi-\theta$ & $\vartheta$ & $d$ & $N$\tabularnewline
\hline
 & Unit & {[}\textmu{}V/m/MHz] & - & - & {[}ns] & - & {[}eV] & {[}$\unit{^{\circ}}$] & {[}$\unit{^{\circ}}$] & {[}m] & -\tabularnewline
\hline
\hline 
1 & 2006.03.14.13:30:27 & 11\textpm{}4 & -1.8\textpm{}0.5 & -0.031\textpm{}0.008 & 50 & 6 & (...) & 0-20 & 95 & 70\textpm{}50 & 24\tabularnewline
\hline 
2 & 2006.03.23.04:01:07 & 17\textpm{}4 & -1.1\textpm{}0.4 & -0.019\textpm{}0.007 & 77 & 13 & 3.60E+17 & 35-51 & 60 & 80\textpm{}40 & 23\tabularnewline
\hline
3 & 2006.04.05.05:39:13 & 25\textpm{}5 & -1.1\textpm{}0.8 & -0.02\textpm{}0.01 & 49 & 16 & 5.40E+17 & 49-47 & 61 & 70\textpm{}30 & 23\tabularnewline
\hline
4 & 2006.05.01.10:31:40 & 19\textpm{}4 & -1.9\textpm{}0.4 & -0.031\textpm{}0.006 & 80 & 14 & 1.50E+17 & 286-37 & 64 & 80\textpm{}30 & 23\tabularnewline
\hline
5 & 2006.06.14.11:03:15 & 20\textpm{}6 & -0.4\textpm{}0.3 & -0.008\textpm{}0.005 & 57 & 46 & 1.00E+18 & 340-24 & 89 & 80\textpm{}50 & 14\tabularnewline
\hline
6 & 2006.06.19.19:21:24 & 6\textpm{}3 & -0.8\textpm{}0.3 & -0.013\textpm{}0.004 & 47 & 5 & 1.80E+17 & 12-34 & 79 & 80\textpm{}40 & 20\tabularnewline
\hline
7 & 2006.08.19.23:01:41 & 25\textpm{}6 & -3.6\textpm{}0.4 & -0.061\textpm{}0.006 & 101 & 60 & 2.50E+17 & 20-41 & 73 & 70\textpm{}40 & 9\tabularnewline
\hline
8 & 2006.09.11.05:24:22 & 15\textpm{}5 & -1.2\textpm{}0.4 & -0.021\textpm{}0.006 & 77 & 8 & 1.50E+17 & 341-38 & 75 & 80\textpm{}40 & 9\tabularnewline
\hline
9 & 2006.09.22.17:58:45 & 19\textpm{}6 & -0.8\textpm{}0.6 & -0.01\textpm{}0.01 & 51 & 8 & 4.90E+17 & 56-50 & 57 & 90\textpm{}40 & 8\tabularnewline
\hline
10 & 2006.10.13.10:43:03 & 11\textpm{}4 & -1.9\textpm{}0.8 & -0.03\textpm{}0.01 & 85 & 35 & 1.90E+17 & 0-64 & 51 & 90\textpm{}50 & 9\tabularnewline
\hline
11 & 2006.11.02.07:14:52 & 24\textpm{}7 & -2\textpm{}0.4 & -0.034\textpm{}0.006 & 67 & 9 & 2.20E+17 & 328-30 & 81 & 70\textpm{}40 & 9\tabularnewline
\hline
12 & 2005.12.05.06:46:13 & 29\textpm{}7 & -0.8\textpm{}0.3 & -0.014\textpm{}0.004 & 55 & 13 & 3.10E+18 & 315-22 & 87 & 60\textpm{}30 & 16\tabularnewline
\hline
13 & 2005.12.08.09:10:45 & 31\textpm{}9 & -0.4\textpm{}0.6 & -0.005\textpm{}0.01 & 113 & 10 & (...) & 75-15 & 82 & 60\textpm{}30 & 16\tabularnewline
\hline
14 & 2006.01.18.13:46:06 & 11\textpm{}3 & -1.2\textpm{}0.5 & -0.022\textpm{}0.008 & 59 & 8 & 2.00E+17 & 357-49 & 66 & 70\textpm{}30 & 18\tabularnewline
\hline
15 & 2006.01.19.06:26:13 & 11\textpm{}3 & -1.3\textpm{}0.7 & -0.02\textpm{}0.01 & 52 & 8 & 1.30E+17 & 347-43 & 71 & 80\textpm{}40 & 18\tabularnewline
\hline
16 & 2006.02.04.13:33:28 & 14\textpm{}4 & -0.1\textpm{}0.4 & -0.002\textpm{}0.007 & 51 & 57 & 3.30E+17 & 4-44 & 71 & 70\textpm{}40 & 18\tabularnewline
\hline
17 & 2006.02.07.09:40:20 & 27\textpm{}7 & -1.7\textpm{}0.3 & -0.029\textpm{}0.005 & 49 & 6 & 2.90E+17 & 325-30 & 81 & 100\textpm{}50 & 21\tabularnewline
\hline
18 & 2006.02.10.08:11:20 & 14\textpm{}4 & -0.8\textpm{}0.4 & -0.014\textpm{}0.006 & 52 & 13 & 1.60E+17 & 345-36 & 77 & 80\textpm{}30 & 24\tabularnewline
\hline
19 & 2006.02.14.02:09:17 & 40\textpm{}10 & -1.2\textpm{}0.3 & -0.02\textpm{}0.005 & 56 & 9 & (...) & 28-12 & 101 & 60\textpm{}40 & 16\tabularnewline
\hline
20 & 2006.02.18.19:30:41 & 66\textpm{}10 & -1\textpm{}0.3 & -0.018\textpm{}0.005 & 48 & 292 & 3.60E+17 & 9-37 & 78 & 80\textpm{}40 & 24\tabularnewline
\hline
21 & 2006.02.20.12:26:29 & 40\textpm{}8 & -1.1\textpm{}0.4 & -0.019\textpm{}0.006 & 98 & 88 & 7.70E+17 & 278-26 & 70 & 60\textpm{}40 & 24\tabularnewline
\hline
22 & 2006.02.23.03:34:39 & 9\textpm{}3 & -0.8\textpm{}0.5 & -0.014\textpm{}0.008 & 57 & 7 & 2.10E+17 & 322-43 & 68 & 90\textpm{}40 & 24\tabularnewline
\hline
23 & 2006.03.07.16:55:47 & 11\textpm{}3 & -1.4\textpm{}0.7 & -0.02\textpm{}0.01 & 124 & 8 & 2.60E+17 & 4-65 & 49 & 100\textpm{}50 & 23\tabularnewline
\hline
\end{tabular}
\par\end{raggedright}
\end{sidewaystable}

\end{document}